\documentclass[useAMS,usenatbib]{mn2e}
 \usepackage{relsize,txfonts}
 \usepackage{graphicx,epsfig,bm}
 \usepackage{times}
 \usepackage{rotate}
 \usepackage{graphics}
 \newif\ifAMStwofonts
 \AMStwofontstrue
 \usepackage{amsfonts, amssymb}
\begin{document}

\title[H-ATLAS: Statistical Properties of Galactic Cirrus]{\textit{Herschel}-ATLAS: Statistical Properties of Galactic Cirrus in the GAMA-9 Hour Science Demonstration Phase Field}

\author[A. Bracco et al.]{A. Bracco$^{1}$, A. Cooray$^{1,2}$, M. Veneziani$^{1,3}$, A. Amblard$^{1}$, P. Serra$^{1}$, J. Wardlow$^{1}$, \newauthor  
M.~A. Thompson$^{4}$,  G. White$^{5,6}$, R. Auld$^{7}$, M. Baes$^8$, F. Bertoldi$^{9}$, S. Buttiglione$^{10}$, A. Cava$^{11}$, 
\newauthor  
D.~L. Clements$^{12}$, A. Dariush$^{7}$, G. De Zotti$^{10,13}$, L. Dunne$^{14}$, S. Dye$^{7}$, S. Eales$^{7}$, J. Fritz$^8$, 
\newauthor
H. Gomez$^{7}$, R. Hopwood$^{5}$, I. Ibar$^{15}$,  R.~J. Ivison$^{15,16}$, M. Jarvis$^{4}$, G. Lagache$^{17,18}$, 
\newauthor
M.~G. Lee$^{19}$, L. Leeuw$^{20}$, S. Maddox$^{14}$, M. Micha{\l}owski$^{16}$, C. Pearson$^{21,5,22}$, M. Pohlen$^{7}$, 
\newauthor
E. Rigby$^{14}$,  G. Rodighiero$^{23}$,  D.J.B. Smith$^{14}$,  P.Temi$^{24}$, M. Vaccari$^{23}$, P. van der Werf$^{16,25}$
\\
 {$^{1}$Dept. of Physics \& Astronomy, University of California, Irvine, CA 92697, USA}\\
 {$^{2}$Division of Physics, Math \& Astronomy, California Institute of Technology, Pasadena, CA 91125, USA}\\
 {$^{3}$Spitzer Science Center, California Institute of Technology, Pasadena, CA 91125, USA}\\
 {$^{4}$Centre for Astrophysics Research, Science and Technology Research Institute, University of Hertfordshire, Herts AL10 9AB, UK}\\
 {$^5$Department of Physics \& Astronomy, The Open University, Walton Hall, Milton Keynes MK7 6AA, UK}\\
 {$^6$Space Science \& Technology Division, The Rutherford Appleton Laboratory, Chilton, Oxfordshire OX11 0NL, UK}\\
 {$^{7}$School of Physics and Astronomy, Cardiff University, The Parade, Cardiff, CF24 3AA, UK}\\
 {$^{8}$Sterrenkundig Observatorium, Universiteit Gent, Krijgslaan 281 S9,B-9000 Gent, Belgium}\\
 {$^{9}$INAF - Osservatorio Astronomico di Padova,  Vicolo Osservatorio 5, I-35122 Padova, Italy}\\
 {$^{10}$Argelander-Institute for Astronomy, University of Bonn, Auf dem Huegel 71, D-53121 Bonn, Germany}\\
 {$^{11}$Instituto de Astrofısica de Canarias and Departamento de Astrofisica - Universidad de La Laguna, E38205, La Laguna, Spain}\\
 {$^{12}$Astrophysics Group, Physics Department, Imperial College, Prince Consort Road, London, SW7 2AZ, UK}\\
 {$^{13}$SISSA, Via Bonomea 265, I-34136 Trieste, Italy}\\
 {$^{14}$School of Physics and Astronomy, University of Nottingham, University Park, Nottingham NG7 2RD, UK}\\
 {$^{15}$UK Astronomy Technology Centre, Royal Observatory, Edinburgh, EH9 3HJ, UK}\\
 {$^{16}$SUPA, Institute for Astronomy, University of Edinburgh, Royal Observatory, Blackford Hill, Edinburgh, EH9 3HJ, UK}\\
 {$^{17}$Univ Paris-Sud, Laboratoire IAS, UMR8617, Orsay, F-91405, France}\\
 {$^{18}$CNRS, Orsay, F-91405, France}\\
 {$^{19}$Department of Physics and Astronomy, Seoul National University, Seoul 151-742, Korea}\\
 {$^{20}$SETI Institute, 515 N. Whisman Avenue, Mountain View, CA, 94043}\\
 {$^{21}$Space Science \& Technology Department, Rutherford Appleton Laboratory, Oxon, OX11 0QX, UK}\\
 {$^{22}$Institute for Space Imaging Science, University of Lethbridge, Lethbridge, Alberta T1K 3M4, Canada}\\
 {$^{23}$Department of Astronomy, University of Padova, Vicolo Osservatorio 3, Padova, Italy}\\
 {$^{24}$Astrophysics Branch, NASA Ames Research Center, Mail Stop 245-6, Moffett Field, CA 94035, USA}\\
 {$^{25}$Leiden Observatory, Leiden University, P.O. Box 9513, NL - 2300 RA Leiden, The Netherlands}\\
}
\voffset=-0.8in


\maketitle
\def\mnras{MNRAS}

\begin{abstract}
We study the Spectral Energy Distribution (SED) and the power spectrum of Galactic cirrus 
emission observed in the 14 deg$^2$ Science Demonstration Phase field of the {\it Herschel}-ATLAS  using {\it Herschel} and {\it IRAS} data from 100 to 500$\,\mu$m. We compare the SPIRE 250, 350 and 500\,$\mu$m maps with {\it IRAS} 100\,$\mu$m emission, binned in $6^{\prime}$ pixels.  We assume a modified black-body SED with dust emissivity parameter $\beta$ ($F \propto \lambda^{-\beta}$) and a single dust temperature $T_{\rm d}$, and find that the dust temperature and emissivity index varies over the science demonstration field as $10< T_{\rm d} < 25\, \rm K$ and $1< \beta< 4$.   The latter values are somewhat higher than the range of $\beta$ often quoted in the literature ($1< \beta< 2$). We estimate the mean values of these parameters to be $T_{\rm d}=19.0  \pm 2.4\,{\rm K}$ and $\beta  = 1.4 \pm 0.4$.    In regions of bright cirrus emission, we find that the dust has similar temperatures with $T_{\rm d}=18.0 \pm 2.5\rm \,{\rm K}$, and similar values of $\beta$, ranging from $1.4\pm 0.5$ to $1.9\pm 0.5$.    We show that $T_{\rm d}$ and $\beta$ associated with diffuse cirrus emission are anti-correlated and can be described by the relationship: $\beta(T_{\rm d}) = NT_{\rm d}^{\alpha}$ with [$N=116 \pm 38$, $\alpha=-1.4 \pm 0.1$].  The strong correlation found in this analysis is not just limited to high density clumps of cirrus emission as seen in previous studies, but is also seen in diffuse cirrus in low density regions.  To provide an independent measure of  $T_{\rm d}$ and $\beta$, we obtain the angular power spectrum of the cirrus emission in the {\it IRAS} and SPIRE maps, which is  consistent with a power spectrum of the form $P(k)=P_0(k/k_0)^{\gamma}$ where $\gamma = −2.6 \pm 0.2$ for scales of $50-200^{\prime}$ in the SPIRE maps. The cirrus {\it rms} fluctuation amplitude at angular scales of $100^{\prime}$ is consistent with a modified blackbody SED with $T_{\rm d} = 20.1\pm 0.9\,\rm K$ and $\beta = 1.3\pm 0.2$, in agreement with the values obtained above.
\end{abstract}
\section{Introduction}

The sub-millimeter and millimeter emission of diffuse Galactic dust is primarily determined 
by the thermal radiation of large dust grains that are in equilibrium with the  interstellar radiation field 
(D\'esert et al 1990).  The dust organizes itself into large-scale structures such as cirrus and filaments 
both at low and high Galactic latitudes (Low 1984). To compare with previous studies we approximate the dust emission
by a single thermal Planck spectrum at the 
temperature $T_{\rm d}$ of the grains modified by a power law dependence on frequency parameterized by
the spectral emissivity parameter $\beta$ in the optically thin approximation:
\begin{equation}
I(\nu) = \epsilon(\nu)B(\nu, T_{\rm d}) N_{\rm H}\, ,
\end{equation}
where $I(\nu)$ is the specific brightness, $B_\nu$ is the Planck spectrum, and $N_H$ 
is the total hydrogen column density along the line of sight. The isothermal assumption is likely to be only approximate.
While at high latitudes the overlap of several dust sources along the line of sight is expected to be small, we still expect a complex temperature
structure due to variations in the grain size distribution and any variations in the radiation field.
In the above equation, $\epsilon(\nu)$
is the emissivity
\begin{equation}
\epsilon(\nu) = X_d\epsilon_0\left(\frac{\nu}{\nu_0}\right)^\beta \,
\end{equation}
where $X_d$ is the dust-to-gas mass ratio and $\epsilon_0$ is the emissivity at frequency $\nu_0$. 

\begin{figure*}
\begin{center}
\includegraphics[scale=0.98]{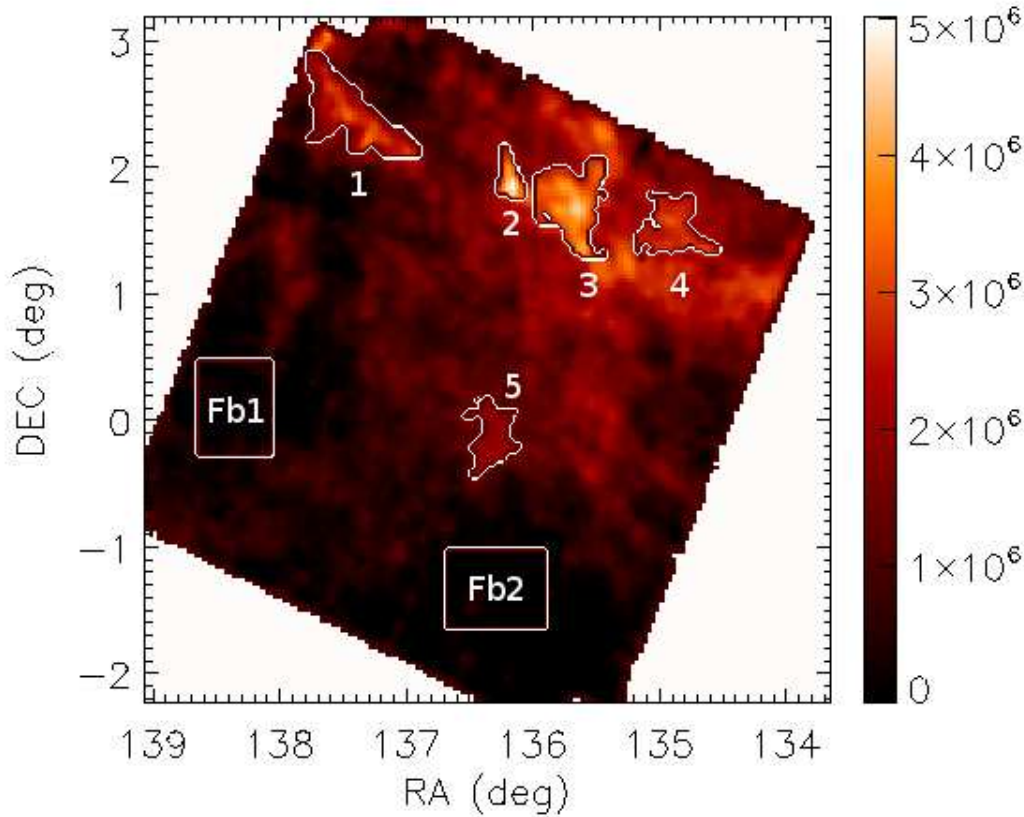}
\end{center}
\caption{Selected regions of  high cirrus intensity that we have individually analyzed (labeled 1 to 5)
as well as the two low-cirrus regions (labeled Fb1 and Fb2) that we used to account for the extragalactic background intensity. The map, at 250 $\mu$m, is color coded in intensity units of Jy/sr.}
\label{fig:reg1}
\end{figure*}

With $T_{\rm d}$ at the level of 10 to 30K the maximum intensity is found at far-infrared/sub-mm wavelengths.
Due to the lack of coverage at far-IR wavelengths, studies on the Galactic cirrus temperature and emissivity 
before {\em Herschel}\footnote{{\em Herschel} is an ESA space observatory with science instruments 
provided by European-led Principal Investigator consortia and with important participation from NASA.} (Pilbratt et al. 2010) focused on
the wavelength bands shorter than 160 $\mu$m that were covered by {\it IRAS} and {\it Spitzer} (e.g., Miville-Desch\^enes et al. 2002; Jeong et al. 2005),
longer than 1 mm covered by various cosmic microwave background (CMB) experiments (e.g., D\'esert et al. 2009; Veneziani et al. 2010),
and a combination including limited sub-mm data (e.g., Dupac et al. 2003; Paradis et al. 2009; Bernard et al. 2009).

With its unprecedented angular resolution and the ability to cover wide fields by scanning across the sky, 
 the Spectral and Photometric Imaging Receiver (SPIRE; Griffin et al. 2010) on
{\em Herschel} now allows for the first time the possibility to study the structure of the ISM at tens of arcseconds 
 to degree angular scales  at the peak of the dust spectral energy distribution (SED). At the smallest angular scales probed by SPIRE
the structure of the diffuse interstellar medium  provides information about
 the initial conditions for the formation of dense molecular clouds. Some of these clouds may be on the verge of  gravitational collapse leading to the formation
of a new star (Olmi et al. 2010, Sadavoy et al. 2010, Ward-Thompson et al. 2010). At large angular scales,
dust  is a key tracer of the large-scale physical processes occurring in the diffuse interstellar medium (Miville-Desch\^enes et al. 2010).
Dust emission is  also related to the density structure of diffuse  clouds and could potentially provide a way to study the projected 
density distribution within the Galactic cirrus.

In equation~(2), the dust emissivity index $\beta$ provides information on the physical nature of dust and connects the grain structure to the large-scale environmental density. The spectral index of the emissivity depends on grain composition, temperature distribution of tunneling states, and the
wavelength-dependent excitation (e.g., Meny et al. 2007). The emissivity $\beta$  is also expected to vary with wavelength
when the dust temperature is intrinsically multi-component but described by an isothermal model (Paradis et al. 2009).  While the dust SED models 
in the literature generally assume a fixed value for the spectral emissivity $\beta$  between 1.5 and 2.5, there might be significant variations in $\beta$, even in a small
cirrus region, when taking into account the disordered structure of dust grains. 
There are also  however the uncertainties associated with the dust size distribution and the silicate versus graphite fractions; large variations in both these quantities could result in
different equilibrium temperatures even in a small cirrus region. This complicates any physical interpretation of $\beta$ and $T_{\rm d}$ 
when using an isothermal SED. Instead of multi-component models, we use the isothermal model here so that we can compare our results to
 previous analyses that make the same assumption. Our SED modeling is also limited to four data points.
Of particular interest to this work  is the suggestion that $\beta$ and $T_{\rm d}$ are inversely related
in high density environments of Galactic dust by previous observations in the sub-millimeter and millimeter domain, both at low (Dupac et al 2003, D\'esert et al 2009) 
and high (Veneziani et al 2010) Galactic latitudes.

Most of the recent studies on properties of the Galactic cirrus focused on high density environments,
such as cold clumps and molecular clouds  with intensities of order 100 MJy sr$^{-1}$ or more at far-IR wavelengths (see, however, Bot et al. 2009 for a study on diffuse
medium at small scales). Properties of the  interstellar medium at high latitudes, especially involving diffuse cirrus with intensities of order a few MJy sr$^{-1}$, are
still not well known. A good modeling of diffuse dust distribution and its characteristics in the high latitude regions 
is necessary in order to remove its contamination from CMB anisotropy measurements (see for example Leach et al. 2008 and Ricciardi et al. 2010). 
The CMB community primarily relies on models developed with {\it IRAS}  and DIRBE maps to describe the dust distribution 
(e.g., Schlegel et al. 1998) and the frequency dependence of the intensity  (e.g., Finkbeiner et al. 1999).
With wide-field {\em Herschel}-ATLAS (H-ATLAS; Eales et al. 2010) maps we can study the dust temperature and emissivity variation across large areas on the sky,
first in the SDP field covering 14 deg$^2$ and eventually over 550 deg$^2$ spread over 5 fields with varying Galactic latitudes.
The H-ATLAS SDP  patch is at a Galactic latitude of $\sim$ 30 degrees.                                                                                           
Combining SPIRE data at 250, 350, and 500 $\mu$m with {\it IRAS} maps of the same area at 100 $\mu$m allows us to sample
the peak of the dust SED accurately. 

Here we present an analysis of diffuse Galactic cirrus in the H-ATLAS SDP field from 100 $\mu$m to 500 $\mu$m using {\it IRAS} and {\it Herschel}-SPIRE maps. 
We derive physical parameters of the diffuse dust at arcminute angular scales such as the temperature and spectral emissivity parameter and 
their relationship to each other. We also present a power spectrum analysis of the cirrus emission,
which allows us to study the spatial structure of the interstellar medium from tens of arcseconds to degree angular scales. 
The discussion is organized as follows: Section 2 describes the datasets; Section 3 describes the pipeline adopted. Section~4 reports results related to the temperature and
spectral emissivity parameter while Section 5 describes the cirrus power spectrum.  We conclude with a summary in Section 6.


\section{Data Sets}
\label{sec:datasets}

We use {\em Herschel}-SPIRE maps in the  H-ATLAS 14 deg.$^2$ Science Demonstration Phase (SDP) field, centered at
RA=$9^{\rm h}25^{\rm m}31^{\rm m}$, DEC=$0^{\rm d}29^{\rm m}58^{\rm s}$, overlapping with the
GAMA survey (Driver et al. 2009). In addition to the three SPIRE bands, we also use the
{\it IRAS} 100 $\mu$m map. The latter is  obtained by IDL routines projecting the Healpix\footnote{http://healpix.jpl.nasa.gov} format map made by the 
IRIS processing system of the {\it IRAS} survey\footnote{http://www.cita.utoronto.ca/~mamd/IRIS/} (Miville-Desch\^enes \& Lagache 2005). 

We refer the reader to Pascale et al. (2010) for details on the H-ATLAS SPIRE map making procedure and basic details related to the maps. 
Since we are interested in the diffuse emission, we make use of a set of maps
that have been especially made to preserve the extended structure by accounting for the map-making transfer function. We also produced a second set of maps using 
the same timelines processed by HIPE (Ott et al. 2006), but with an independent map-making pipeline. This involved the use of
an iterative approach to make new maps using SHIM v1.0 (The SPIRE-HerMES Iterative Mapper; Levenson et al. 2010). For that map maker 
 simulations show a transfer function that is close  to unity over arcminute to degree angular scales. 
 The results we describe here, however, are consistent within overall uncertainties between the two sets of maps. Thus, we describe results primarily using the
H-ATLAS map making pipeline of Pascale et al. (2010). We do not use the H-ATLAS PACS SDP maps for this analysis since the diffuse emission at short wavelengths imaged by PACS is
heavily filtered out during the map-making process as carried out in the production of H-ATLAS PACS SDP data products (Ibar et al. 2010). 

\begin{figure}
\includegraphics[scale=0.5]{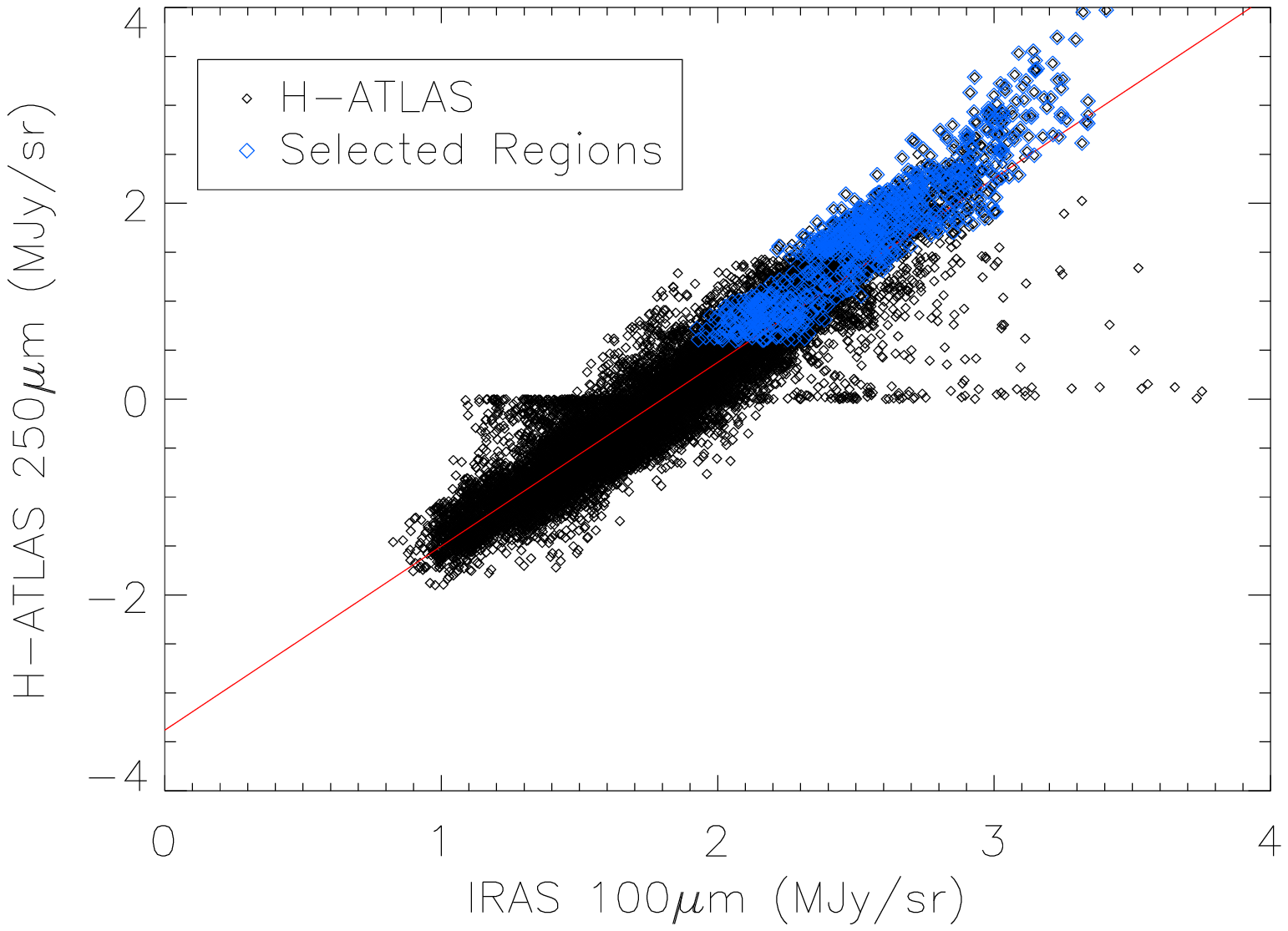}
\includegraphics[scale=0.5]{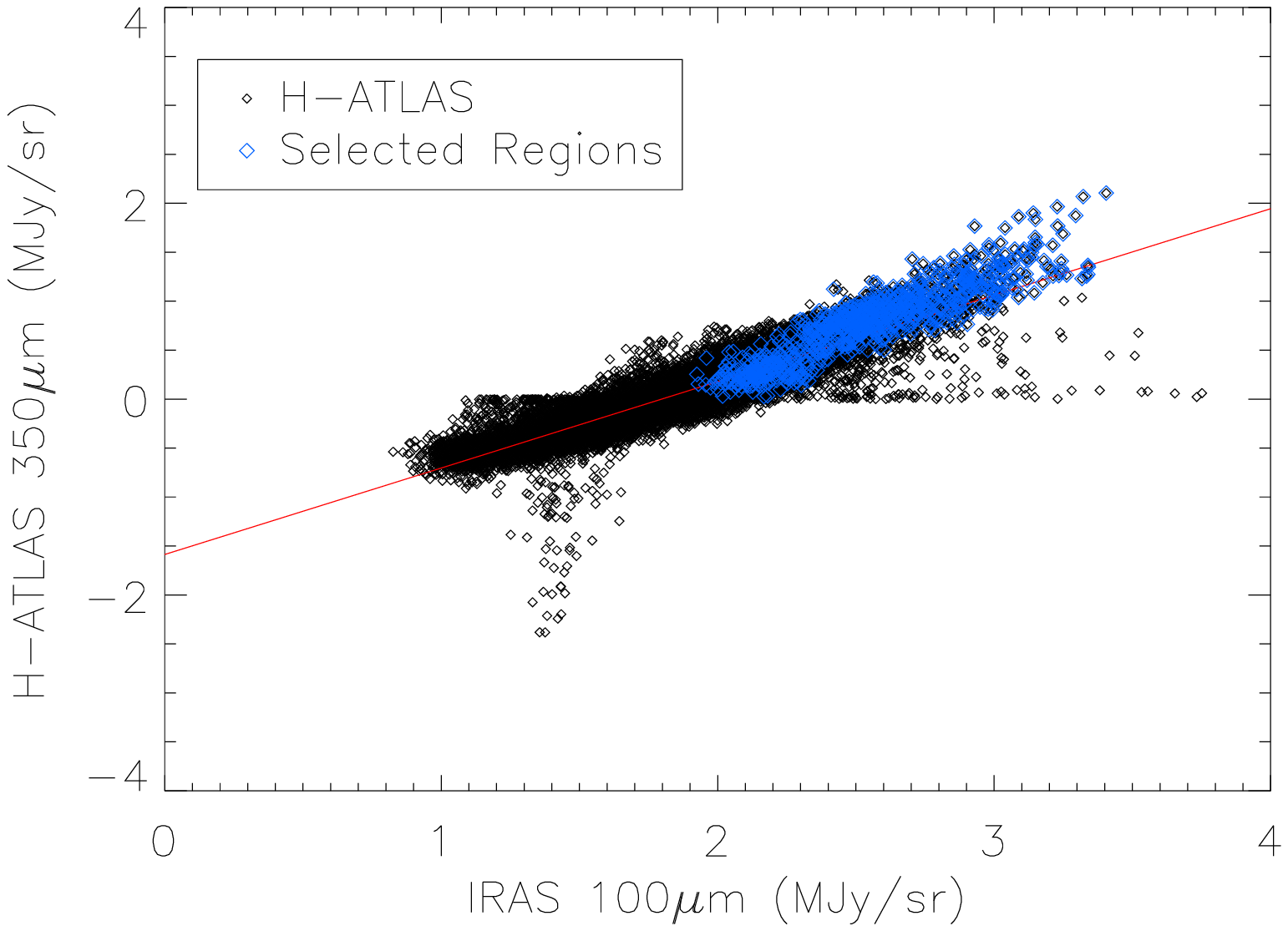}
\includegraphics[scale=0.5]{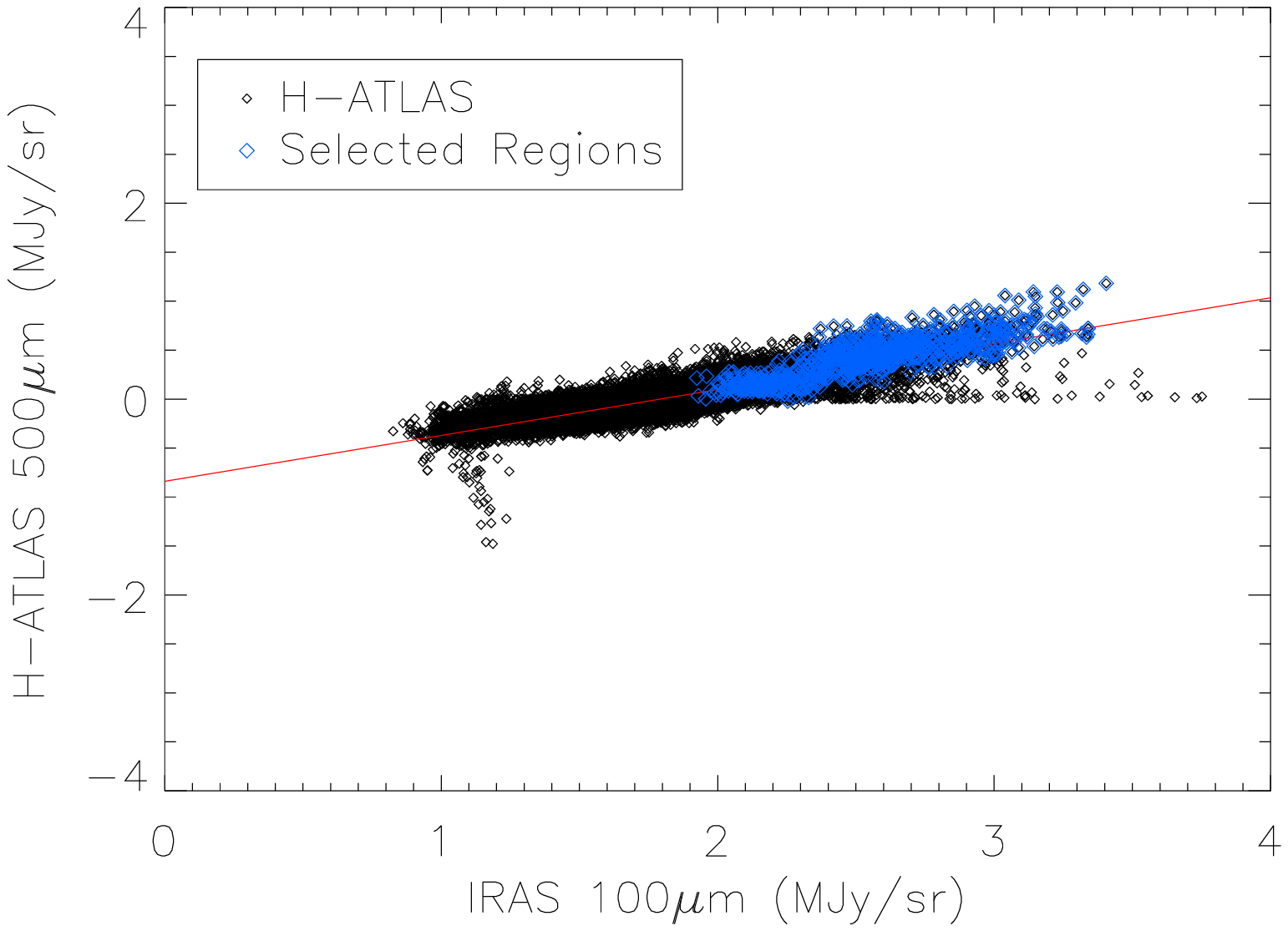}
\caption{The {\it IRAS} 100 $\mu$m intensity vs. SPIRE intensity at 250 (top), 350 (middle), and 500 (bottom) $\mu$m. The SPIRE maps are
pixelized to {\it IRAS} and we use the relation between SPIRE intensity and {\it IRAS} intensity to determine the  sub-mm color  
and zero-point offset relative to
the {\it IRAS} map (see text for details). The points colored blue are from the bright cirrus regions marked 1 to 5 in Figure~1.
The points with zero values to SPIRE intensity are pixels that remain zero after smoothing to {\it IRAS} pixel scale 
due to our point and extended source masks that were defined in the original pixel scale. These and other points that fall off the main band of points, possibly associated with
issues related to parts of the time-streams that are either contaminated or contain glitches that were removed, do not bias the color 
and zero-point offset estimates.}
\label{fig:spireiras}
\end{figure}

\begin{table*}
\begin{center}
\caption{Dust intensity and SED properties of selected high-intensity features from the SDP map}
\label{table:data}      
\centering          
\begin{tabular}{l l l l l l}
\hline\hline
Region $\#$ & 1 &  2 & 3 & 4 & 5 \\ 
\hline
RA (deg) & 137.44 & 136.18  & 135.69  & 134.79 & 136.35  \\
DEC (deg) & 2.39 & 1.90  & 1.63  & 1.47 & -0.31  \\
Area (deg$^2$) & 0.33 & 0.063  & 0.29  & 0.18 & 0.14  \\
\hline
Flux 100 $\mu$m (MJy sr$^{-1}$) & 1.2$\pm$0.1   & 1.5$\pm$0.1  & 1.6$\pm$0.1  & 1.2$\pm$0.1  & 0.9$\pm$0.1 \\
Flux 250 $\mu$m (MJy sr$^{-1}$) & 2.4$\pm$0.4   & 3.5$\pm$0.5  & 3.3$\pm$0.5  & 2.7$\pm$0.4  & 1.8$\pm$0.3  \\
Flux 350 $\mu$m (MJy sr$^{-1}$) & 1.1$\pm$0.2   & 1.8$\pm$0.3  & 1.6$\pm$0.3  & 1.3$\pm$0.2  & 0.7$\pm$0.1   \\
Flux 500 $\mu$m (MJy sr$^{-1}$) & 0.5$\pm$0.1 & 0.9$\pm$0.1  & 0.8$\pm$0.1  & 0.7$\pm$0.1  & 0.4$\pm$0.1  \\
\hline      
ln A &  $-9.1\pm$0.9 & $-9.2\pm$0.8 & $-9.0\pm$0.9  & $-9.5\pm$0.9 & $-9.3\pm$1.1   \\
$\beta$ & 1.8$\pm$0.5 & 1.4$\pm$0.4 & 1.6$\pm$0.5 & 1.4$\pm$0.5 & 1.9 $\pm$ 0.6\\
$T_{\rm d}$ & 17.6$\pm$2.3 & 18.3$\pm$2.2  &  18.1$\pm$2.5 & 18.3$\pm$2.5  & 17.4$\pm$2.9  \\
$\chi^2$ & 0.8 & 0.5 & 0.6 & 1.1 & 2.7 \\ 
\hline
\end{tabular}  
\end{center}
\end{table*}

\begin{figure}
\includegraphics[scale=0.5]{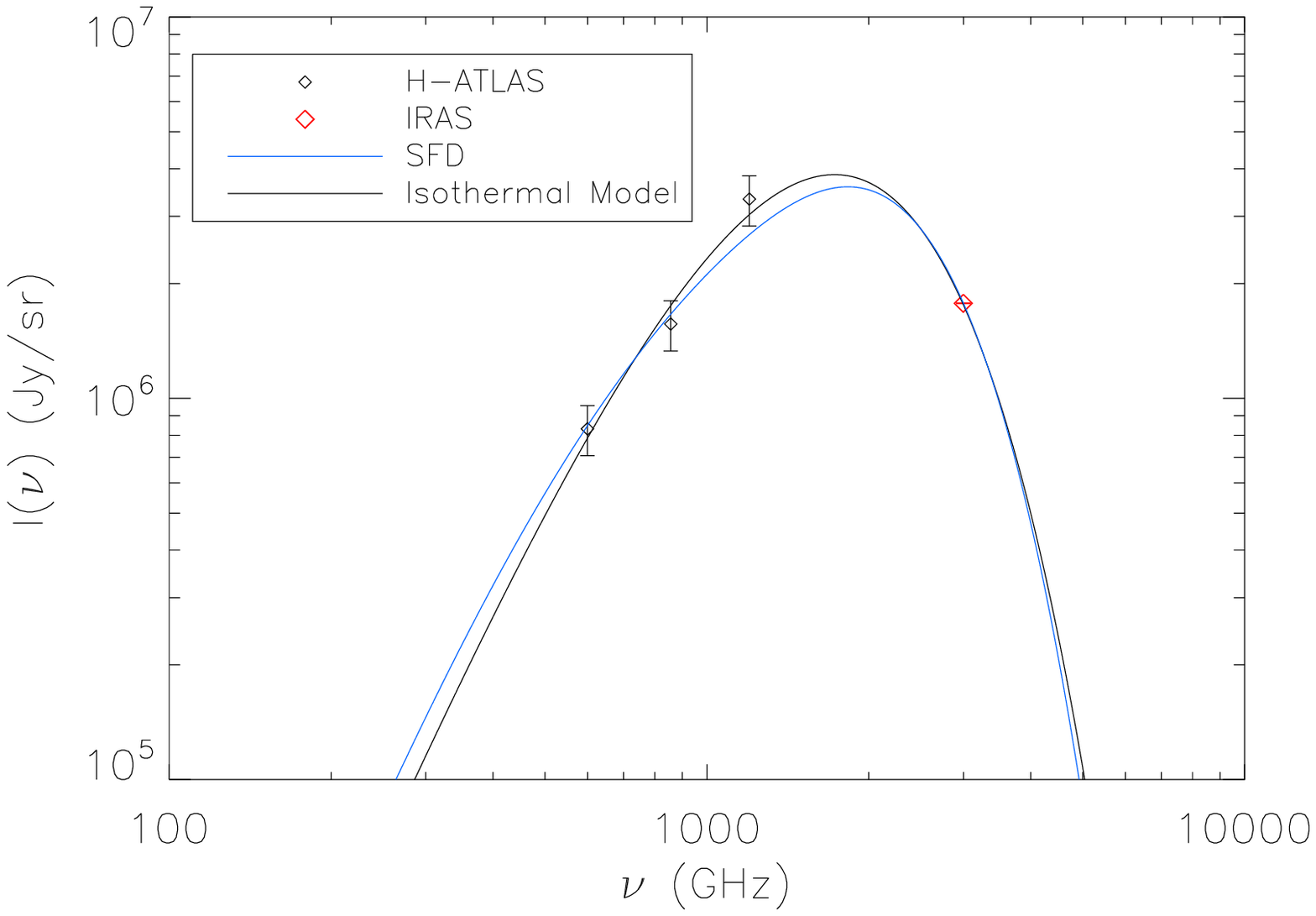}
\caption{The Galactic dust SED averaged over the 14 deg$^2$ H-ATLAS SDP field  and 
constructed by correlating SPIRE pixel intensities with 100 $\mu$m {\it IRAS} pixel intensity to measure the relative sub-mm colors 
at each of the wavelengths (see text for details)
and then scaling the {\it IRAS} mean flux over the field to SPIRE bands. The best-fit dust temperature and spectral emissivity
parameters are $T_{\rm d}=19.0 \pm 2.4$\,K and $\beta=1.4 \pm 0.4$, respectively (black solid line). For reference, we also show the expected SED following  Schlegel et al. (1998) dust map
and Finkbeiner et al. (1999) frequency scaling that involves two temperature components (see Section~4.5; blue solid line).}
\label{fig:allplots}
\end{figure}

\section{Data Analysis}

The aim of this paper is to characterize the physical properties of dust emission over the whole SDP area. 
In order to avoid contamination of our Galactic dust measurements from extragalactic point sources, we first remove the
bright detected sources from each of the maps making use of the H-ATLAS source catalogs (Rigby et al. 2010). 
This catalog involves sources that have been detected at 5$\sigma$ in at least one of the bands. Given that SPIRE data have beam sizes of
approximately 18, 25, and 36$''$, respectively, at 250, 350 and 500 $\mu$m, we introduce a source mask by simply setting the pixel values over a square size of 20, 30, and 40$''$ to be
zero at the source locations. For a handful of extended sources in the Rigby et al. (2010) catalog, we increased the size of the mask based on the source size
as estimated directly from maps. With close to 6700 sources in total, this masking procedure involved a removal of 2, 5 and 6\% of the data at 250, 350 and 500 $\mu$m, respectively.
Removal of such a small fraction of pixels does not bias our results. Alternatively, we could have modeled each source by fitting the PSF at each of the source locations and removing the flux
associated with the source  and retaining the background; we ran a set of simulations to study if the two approaches lead to different results, but we did not find any.
In the case of the 100 $\mu$m {\it IRAS} map,  we similarly masked roughly 35 point sources in the SDP area from the {\it IRAS}-Faint Source Catalog of Wang \& Rowan-Robinson (2009)
by setting the pixel intensity to be zero over a square area of size 6$'$. We also found consistent results when we replace all {\it IRAS} pixels above 5 $\sigma$ with zero intensity.

To compare the {\it IRAS} and {\em Herschel} maps, we then convolve the source-masked SPIRE maps to the angular resolution of the {\it IRAS} 100 $\mu$m map (258$''$ FWHM)  and repixelize 
SPIRE maps at the same pixel scale as {\it IRAS} (120$''$ pixel sizes).    For the {\it IRAS} map, the flux errors are estimated by assuming that noise is isotropic and using the IRIS noise estimate 
(Miville-Desch\^enes \& Lagache 2005).
For SPIRE, we use the noise maps produced by taking the
differences of repeated scans (Pascale et al. 2010) and convolve the noise map to the {\it IRAS} 100 $\mu$m angular resolution and {\it} IRAS pixel size.

\section{Dust Spectral Energy Distribution}

To describe the SED we use a modified black-body spectrum with a spectral emissivity $\beta$ in the optically-thin limit such that:
\begin{equation}
\label{equation:SED}
S(\nu)=A\left(\frac{\nu}{\nu_{0}}\right)^{\beta} B_\nu(T_{\rm d})  
\end{equation}
where the amplitude $A$ depends on the optical depth through the dust,  $\beta$ is the spectral emissivity,
 and $T_{\rm d}$ is the temperature of the dust. We take $\nu_{0}=3$ THz corresponding to the {\it IRAS} 100 $\mu$m measurement.

To estimate the best-fit values for the three unknown parameters $(A,\beta,T_{\rm d})$, we make use of a
Markov Chain Monte Carlo (MCMC) approach (Lewis \& Bridle 2002).  We are able to generate the MCMC chains rapidly
and at the same time fully sample the likelihood functions of the model parameters. Appropriate sampling of the likelihood 
is crucial to study the relation between $\beta$ and $T_{\rm d}$ (Section~4.4).

The model fits adopt uniform priors over a wide range with $\beta$ between -1 and 6 and $T_{\rm d}$ between 1 and 50\,K 
so that we do not incorrectly constrain the best-fit values by a narrow
range of priors. The ranges are set such that we also allow both the minimum and maximum values of $\beta$ and  $T_{\rm d}$ to be  well outside the expected extremes.
We run  MCMC chains for each region until we get convergence based on the Gelman and Rubin statistic (Gelman \& Rubin 1992), with a value for
$1-R$ of at least 0.01 where $R$ is defined as the ratio between the variance of chain means and the mean of the variances.

\subsection{Zero-level in SPIRE maps}

Since the SPIRE maps are not absolutely calibrated, to study the dust temperature over the SDP region as a whole, we
account for the zero-point offset through the pixel-correlation method of Miville-Desch\^enes et al. (2010).
Before computing the correlation we remove a constant intensity corresponding to 
the extragalactic background at 100 $\mu$m from the {\it IRAS} map (0.78 MJy sr$^{-1}$; Lagache et al. 2000; Miville-Desch\^enes et al. 2007). 
We then correlate the {\it IRAS} pixel intensity with that of a SPIRE map at the corresponding pixel. Here we use 
the source-masked SPIRE maps repixelized to the {\it IRAS} pixel scale following the procedure described in Section~3.  

We show this correlation in Figure~2 where we plot the pixel intensity values at the three SPIRE bands as a function of the {\it IRAS} intensity. 
We minimize the difference between $\sum_{ij} [S_{ij}(\lambda)-P_{ij}(\lambda)]$ 
where $S_{ij}(\lambda)$ is the SPIRE intensity at pixel $(i,j)$ at waveband $\lambda$ and 
$P_{ij}(\lambda)$ is the predicted SPIRE flux in each of the pixels by scaling {\it IRAS} 100 $\mu$m map intensity $I(100)$,
with the assumption that $P_{ij}(\lambda)=G\times I_{ij}(100)+S_0$, where $G$, sub-mm color (also called ``gain'' in 
Miville-Desch\^enes et al. 2010), and $S_0$, the zero-point offset, are the mean values over the whole of the map.
This correct for the additive ($S_0$) term under the assumption that the {\it IRAS} map is a true reflection of the sky. 
The estimated values of $G$ are $1.9 \pm 0.3$, $0.9 \pm 0.3$, and $0.5 \pm 0.1$ at 250, 350, and 500 $\mu$m respectively, while $S_0$ takes the values of
$-(3.2 \pm 1.0)$, $-(1.5 \pm 0.6)$, $-(0.8 \pm 0.4)$ MJy sr$^{-1}$ at each of the three frequencies.

The uncertainty in sub-mm color $G$ is estimated by taking the rms of the ratio involving  $S(\lambda)/I(100)$ 
once the offset $S_0$ is removed from each of the maps, while the uncertainty in $S_0$ comes
from the rms of the difference involving $S(\lambda)-G\times I(100)$, with the best-fit values for sub-mm color
 used in the computation. As described in  Miville-Desch\^enes et al. (2010) these rms
values reflect the overall uncertainties more accurately than if one were to simply use the statistics associated with the linear fit to the relation between $S(\lambda)$ and $I(100)$. Those have uncertainties that are at least a factor of 10 smaller than the errors quoted above. 
Figure~2 shows points which show zero fluxes in SPIRE maps or are significantly negative; these are associated with a combination of the source mask
and pixels that were associated with parts of the time-streams that are either contaminated or contain glitches that were removed.
The correlation analysis above accounts for such pixels when degrading the resolution and
we find that such points do not bias the color and zero-point offset estimates we have quoted above.

The mean intensity of the SPIRE maps in the SDP field, once corrected for $S_0$, is around $(4.8,2.5,1.6)$ ($\pm 0.5$) MJy sr$^{-1}$ at 250, 350, and 500 $\mu$m, respectively.
These can be compared to the estimated extragalactic background intensity at each of the three frequencies of 0.85, 0.69, and 0.39 MJy sr$^{-1}$ with an uncertainty at the level of  0.1 MJy sr$^{-1}$ (Fixsen et al. 1998).
The SDP field, on average, is a factor of 4 to 5 brighter than the extragalactic background. The lowest pixel 
intensity values of the SDP field in Figure~2 (once corrected for $S_0$)  allow an independent constraint on the extragalactic background
intensity, but such a study is best attempted in fields where the overall cirrus intensity is similar to or smaller than the expected extragalactic background. 
Fields that span over a wide range of
Galactic longitudes and latitudes are also desirable since such fields allow an additional constraint on determining a constant intensity that is
independent of the location. We will attempt such studies in future works making use of multiple fields in H-ATLAS.

Beyond the mean intensity, the extragalactic background arising from sources below the confusion noise 
has been shown to fluctuate at the few percent level at 30 arcminute angular scales (Amblard et al. 2010). 
Those faint sources are also responsible for roughly 85\% of the  extragalactic background intensity (Clements et al. 2010; Oliver et al. 2010).
We are not able to account for the contamination coming from the unresolved extragalactic background light, but
the fluctuation intensity of 0.1 MJy sr$^{-1}$ at 30 arcminute angular scales 
do not bias the measurements we report here. The background fluctuations
 act as an extra source of uncertainty in our measurements as they introduce
an extra scatter in the intensity measurements from one region to another and that scatter is captured in the overall error budget.

\subsection{Average Dust SED}

Combining the {\it IRAS} 100 $\mu$m average cirrus intensity  over the whole SDP area (1.77 MJy sr$^{-1}$) with the above 
sub-mm color factors, and assuming a 15\% flux uncertainty
coming from the overall calibration of SPIRE (Swinyard et al. 2010), we estimate the
dust temperature $T_{\rm d}$ and $\beta$ to be $19.0 \pm 2.4$\,K and $1.4\pm 0.4$, respectively (Figure~3). 
The same SED can also be described by the two temperature model of Finkbeiner et al. (1999), where we fix $T_{\rm d}=9.2$ and 16.2\,K and $\beta=1.67$ and 2.7, with the SED
 computed using the maps derived from Schlegel et al. (1998) dust map following an analysis similar to the above (see Section~4.5 for more details).
The overall fit, however, has a reduced $\chi^2$ value of 1.4 compared to 0.8 for the isothermal case.
The Polaris flare studied in Miville-Desch\^enes et al. (2010) has a mean intensity of 40 MJy sr$^{-1}$ 
at 250 $\mu$m and $T_{\rm d}=14.5\pm 1.6$\,K, $\beta=2.3 \pm 0.6$ when averaged over the flare.
The SDP field of H-ATLAS is at least a factor of 8 fainter in the mean intensity and has a higher temperature, but a lower value for $\beta$ suggesting that the dust temperature
and $\beta$ vary substantially across the sky depending on the intensity of the dust. 
This complicates simple approaches to Galactic dust modeling with one or two temperatures and $\beta$ values (e.g., Finkbeiner et al. 1999). 
We discuss this further in Section~4.5.

\subsection{SEDs of Bright Cirrus Regions}

Instead of the average $(T_{\rm d},\beta)$ values over the whole field, we also study the SEDs in five bright cirrus regions that are identified 1 to 5 in Figure~\ref{fig:reg1}. 
The previous SED measurement made use of the SPIRE maps that were corrected for sub-mm color, $G$, and $S_0$ 
values obtained by correlating with the {\it IRAS} 100 $\mu$m map
over the whole field and at the pixel scale of IRAS. To study the SED of bright regions,  we now consider a differential measurement  so that we can determine the SED independent of the {\it IRAS}  100 $\mu$m intensity in the SDP field. 
To account for both the extragalactic background and the zero-point, we remove from the three SPIRE maps
the mean intensity from the two low-cirrus regions identified with rectangles in Figure~1. 
These two regions have intensities that are at the low end of SPIRE intensities plotted 
in Figure~2  and, with the previous zero-level included, these intensities are 1.6, 1.2, and 0.9 ($\pm 0.1$) MJy sr$^{-1}$ at 250, 350, and 500 $\mu$m, 
respectively.
We do the same for the {\it IRAS} map and remove the mean intensity of 1.9 ($\pm$ 0.1) MJy sr$^{-1}$ at 100 $\mu$m determined for the same two regions. This is 
necessary to avoid introducing
an unnecessary difference in the relative calibration between SPIRE and IRAS. We account for the uncertainty in this mean removal in our overall error 
budget. 

This procedure allows us to treat the {\it IRAS} intensity independent of SPIRE, but the results we show here do not strongly depend on 
this additional step. When we simply used the {\it IRAS} $S_0$ corrected maps, with SPIRE zero-level fixed to IRAS,
we still recover SEDs that have
 $\beta$ and $T_{\rm d}$  values consistent within uncertainties. The new maps, however, lead to differences in the amplitude due to the overall shift in
the intensity scale. Another way to think about this is that $\beta$ and $T_{\rm d}$ estimates  extracted from the isothermal SED depend on the 
intensity ratios and not the absolute intensity.
The noise is computed by averaging the noise intensity of pixels in the same region as defined for the intensity measurements.
We add quadratically the flux error, the error in the intensity removed from the two low-cirrus regions, and an overall calibration error taken to be 15\% of the intensity
(Swinyard et al. 2010).  The intensity values of the five selected regions are summarized in Table~1. 
We find the dust temperature of  these bright cirrus regions are around $18\pm3$K with $\beta$ around $1.5 \pm 0.5$,
and consistent with results found for large-scale cirrus observations at high Galactic latitudes with a dust temperature of 17.5K (Boulanger et al. 1996).

\begin{figure}
\includegraphics[scale=0.5]{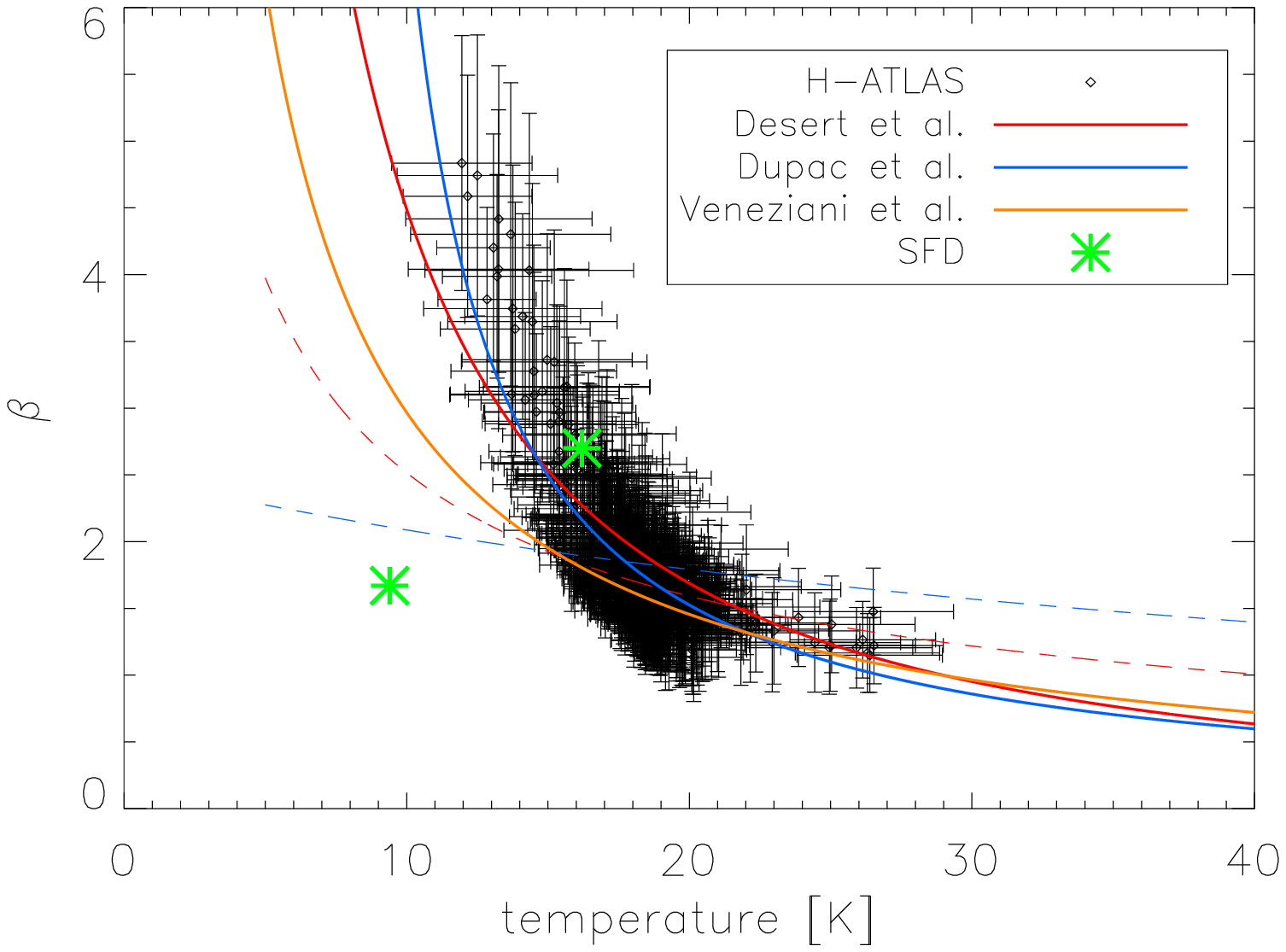}
\includegraphics[scale=0.5]{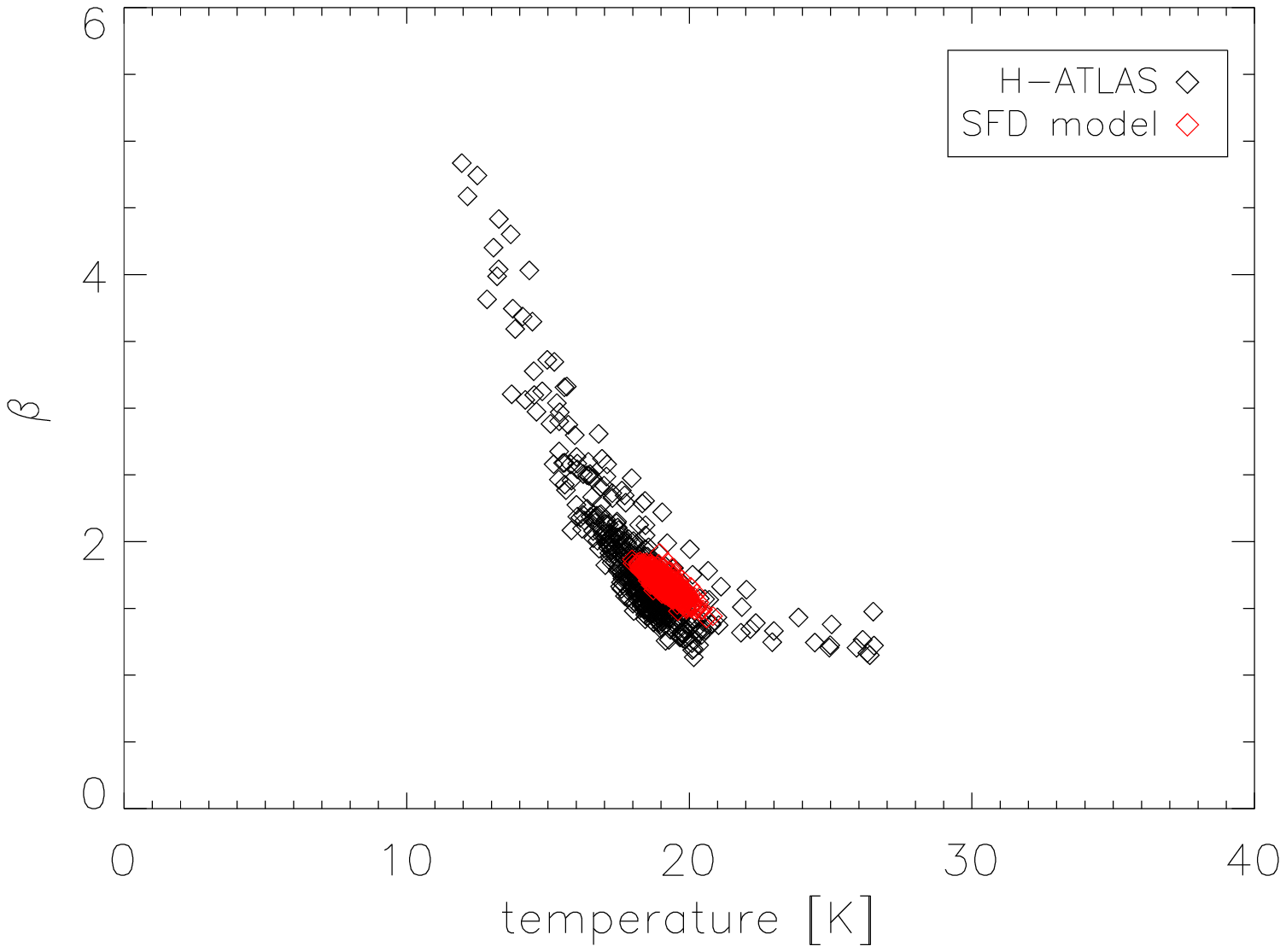}
\caption{{\it Top:} Best-fit values of the spectral emissivity parameter, $\beta$, and dust temperature, $T_{\rm d}$, over the whole SDP field with all maps repixelized at 6$'$ pixels.
The error bars show the 1$\sigma$ uncertainties of these two parameters in each of the pixels. 
The dashed lines show the best-fit models for $T_{\rm d}$ vs. $\beta$ based on the two analytical descriptions outlined in Section~4.4 
with parameter values from Dupac et al. (2003) and D\'esert et al. (2008).  The solid lines show the same model descriptions but with the parameters obtained in this work.  
We also compare the best-fit line of Veneziani et al. (2010) from observations of high density clouds in the BOOMERanG-2003 CMB field.  
Compared to previous measurements that studied high-density dust regions, we find a stronger anti-correlation for diffuse emission.
The noise-weighted mean values of $T_{\rm d}$ and $\beta$ are 20.1 $\pm$ 1.6\, K and 1.5 $\pm$ 0.3, respectively. These are somewhat better than the
previous estimates over the whole area as we exclude pixels which have relative errors for $T_{\rm d}$ and $\beta$ that are greater than 30\%.
The two points show $T_{\rm d}$ and $\beta$ of  Finkbeiner et al. (1999) model-8 used to scale the dust map of Schlegel et al. (1998).
{\it Bottom:} Comparison between best-fit $T_{\rm d}$ and $\beta$ distribution 
with same parameters in the same pixels from the Schlegel et al. (1998)  dust model (see text for details). 
}
\label{fig:corrfit}
\end{figure}

\subsection{Relation between $\beta$ and $T_{d}$}
\label{sec:betatemp}

Beyond the bright regions, we also establish the dust temperature and modified spectral emissivity parameter 
over the whole SDP area. This allows us to produce maps of $T_{\rm d}$ and $\beta$  over the SDP field. To 
this aim we repixelize all maps to $6'\times6'$ pixels resulting in a grid of 55 by 55 pixels for the H-ATLAS SDP field; with smaller pixel sizes
we get noisier estimates of $T_{\rm d}$ and $\beta$ in regions of low cirrus intensity, 
while with larger pixels we do not sample the map adequately. 
The reprojection is done such that the flux is preserved when going from smaller pixels to 6$'$ pixels.
This particular choice of pixel
size  was made so that the total number of grid points used for $(A,T_{\rm d},\beta)$ model fits can be
completed in a reasonable time (a few days in this case) given the computational costs associated with generating separate MCMC chains.
Here again we consider a differential measurement and remove the mean intensity estimated in low-cirrus parts (before the maps were repixelized)
and in those regions we set $A=0$ with the assumption of no cirrus. 

With this analysis we construct the two maps that we show in Figure~5,
where we subselect 6$'$ pixels where the SED fits gave relative errors less than 30\% for both parameters $\beta$ 
and $T_{\rm d}$.  These values are also plotted in Figure~4.
They mostly span the high intensity region of the SDP map.
For reference, In Figure~8 (left panels) we show the maps of $T_{\rm d}$ and $\beta$ without this selection imposed.
At first glance $\beta$ and $T_{\rm d}$ appear to be negatively correlated. Part of this correlation results from the
functional form of the Equation~\ref{equation:SED} that we used to fit the intensity data, especially in the presence of noise (Shetty et al. 2009a). 

In order to discriminate any physical 
effect from the analytical correlation, we test two possible models that describe the $\beta$ versus 
$T_{d}$ dependence, either with
\begin{equation}
\label{equation:desert}
\beta(T_{\rm d})= N T^{\alpha}_{\rm d} 
\end{equation}     
following D\'{e}sert et al. (2008) or
\begin{equation}
\label{equation:dupac}
\beta(T_{\rm d})=\frac{1}{C+x T_{\rm d}}  \, ,
\end{equation}     
from Dupac et al. (2003).

\begin{figure}\label{fig:betatemppixel}
\begin{minipage}{100mm}
\epsfig{file=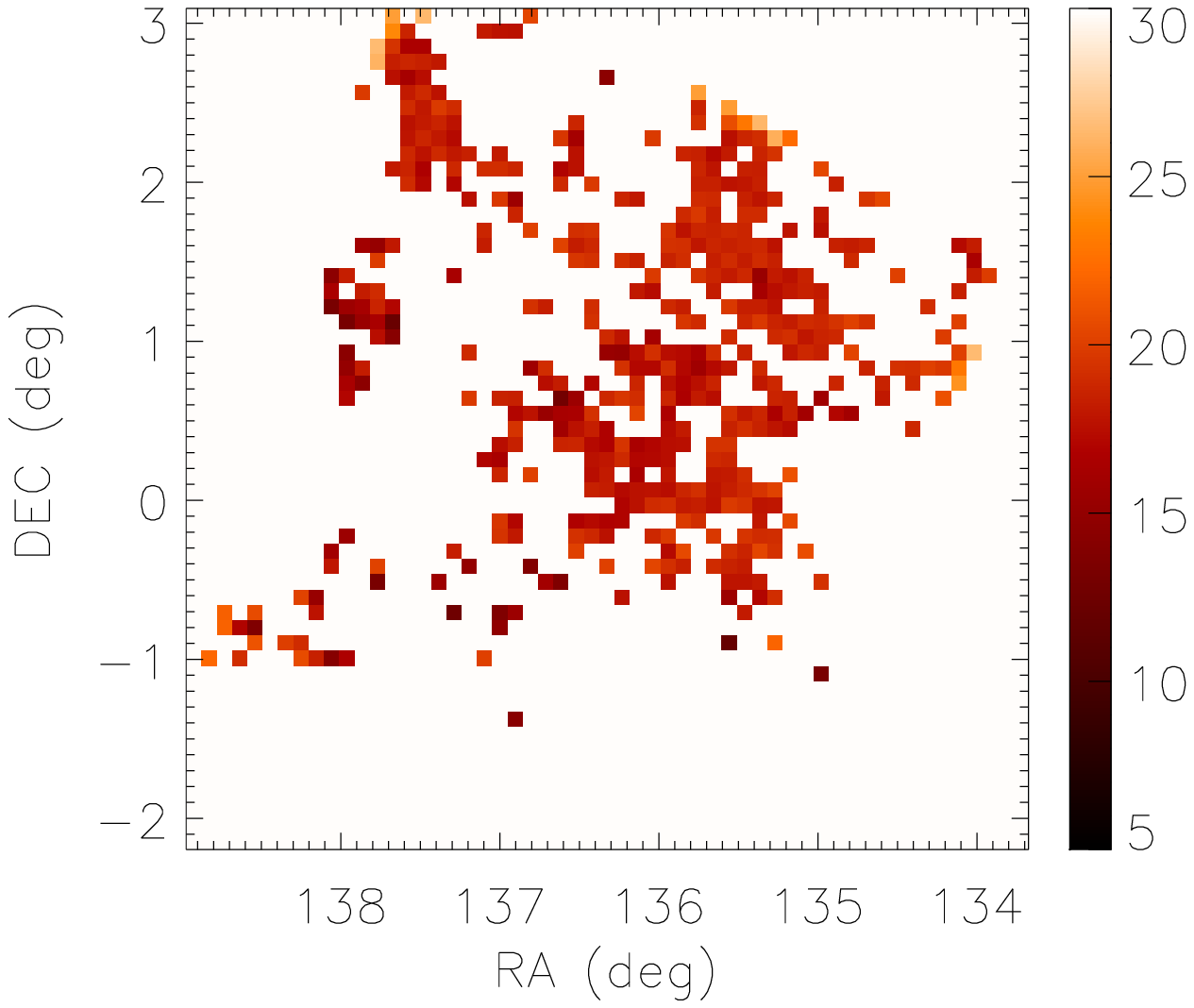,scale=0.6}
\epsfig{file=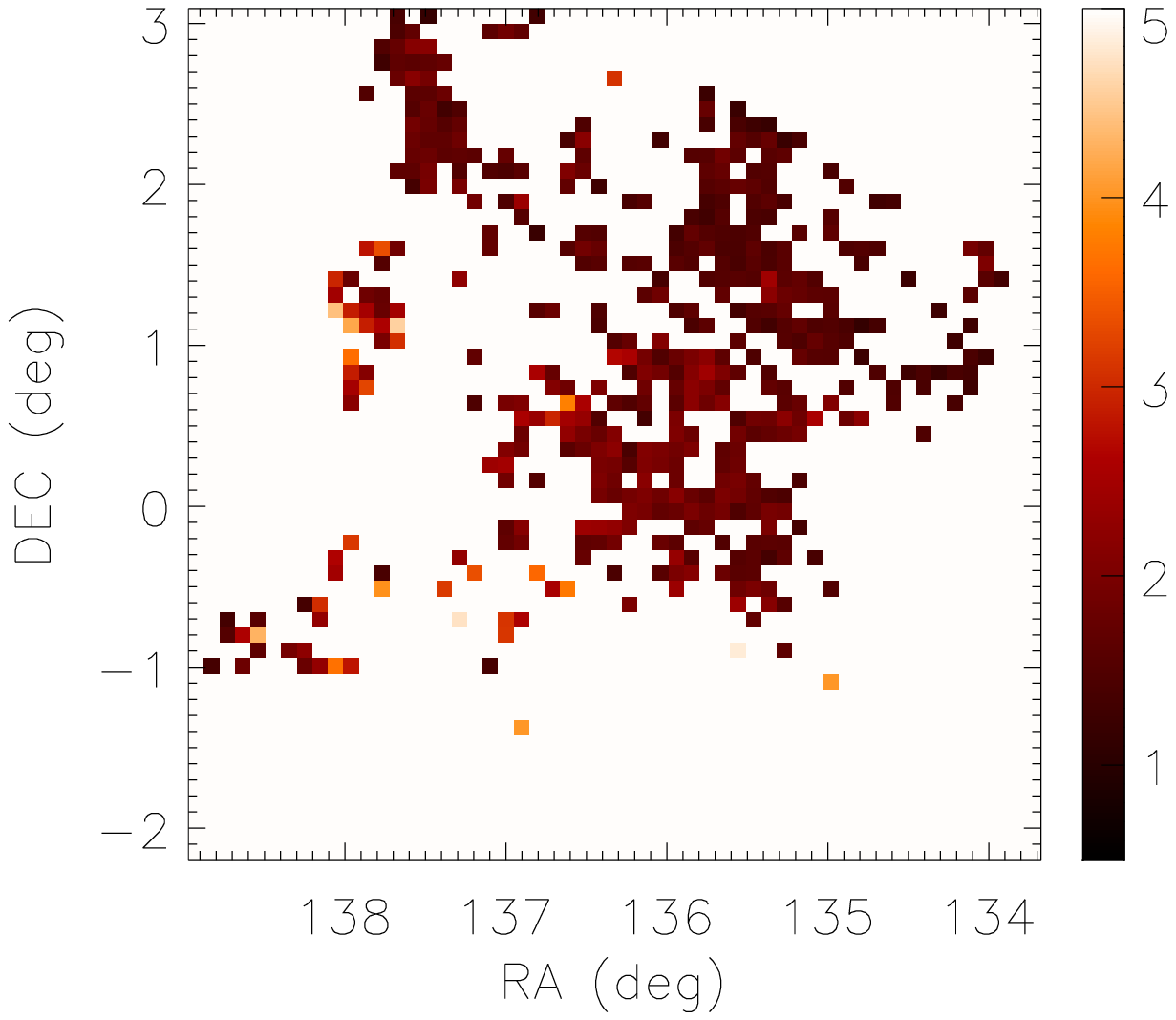,scale=0.6}
\end{minipage}
\caption{The dust temperature (top) and spectral emissivity parameter $\beta$ (bottom) based on our SED fits to pixel intensities that survive the requirement that the two parameters
be measured with relative uncertainty (1$\sigma$) better than 30\%.}
\end{figure}

\begin{figure}\label{fig:exampleSED}
\begin{minipage}{100mm}
\epsfig{file=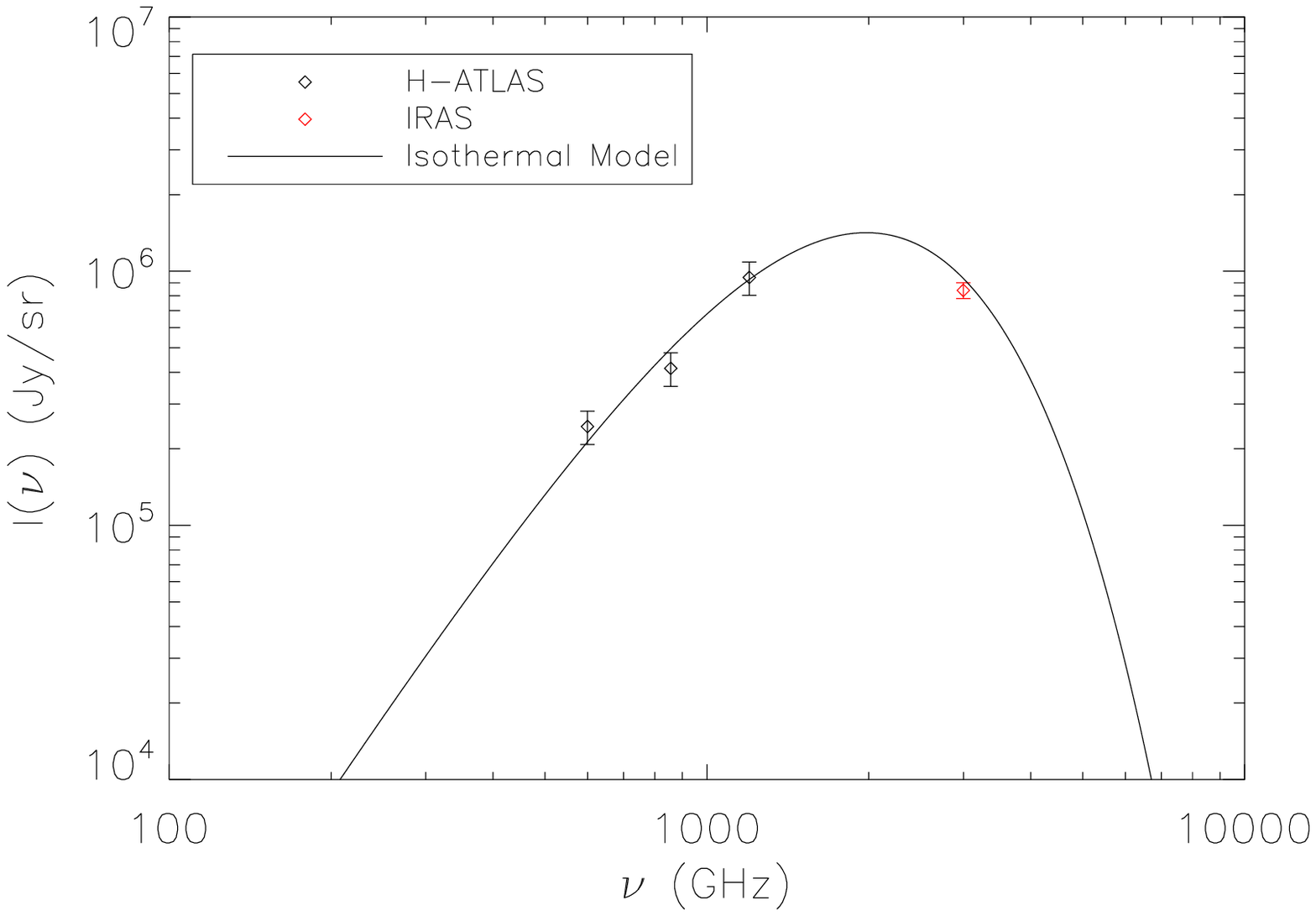,scale=0.45}
\epsfig{file=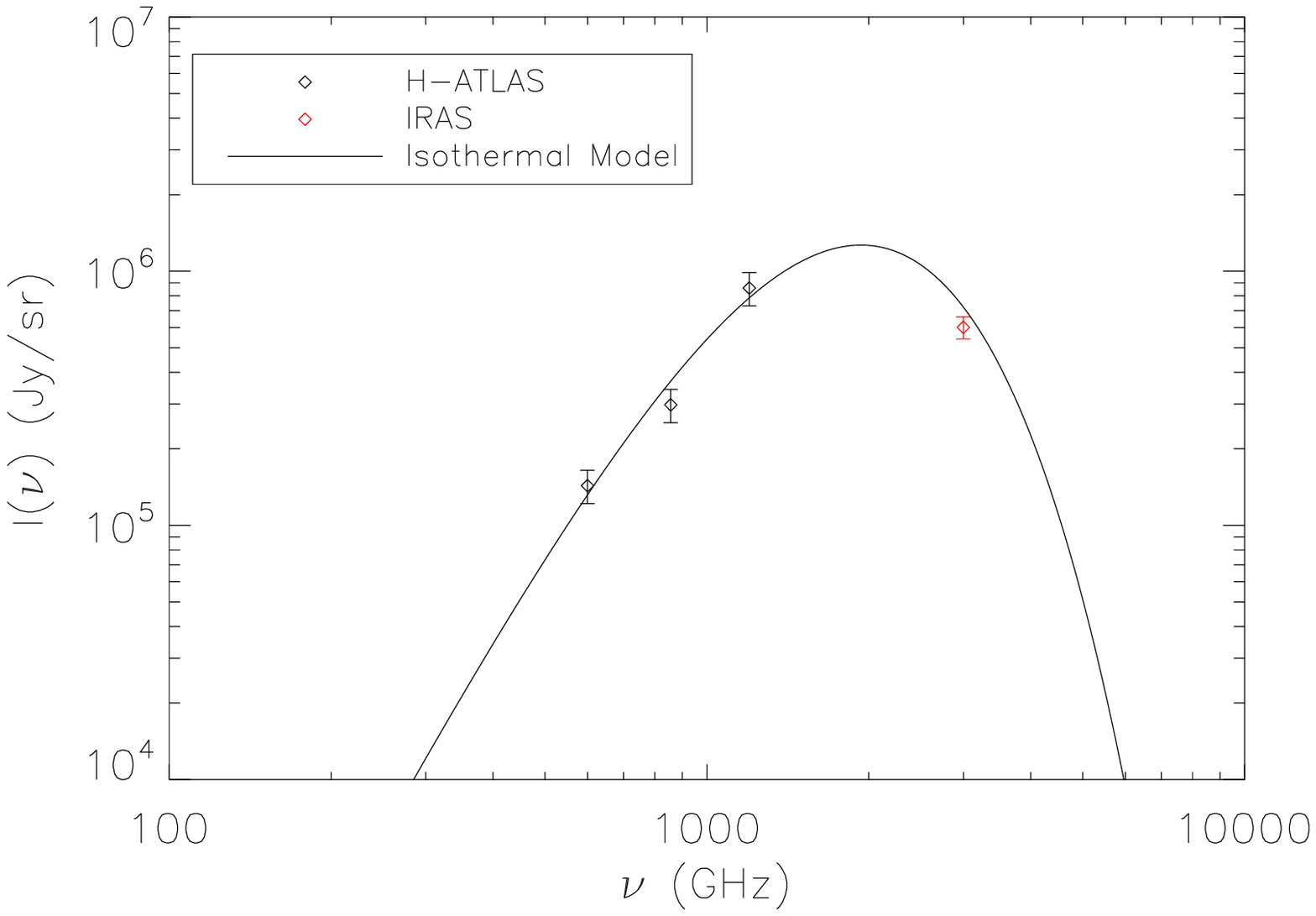,scale=0.45}
\epsfig{file=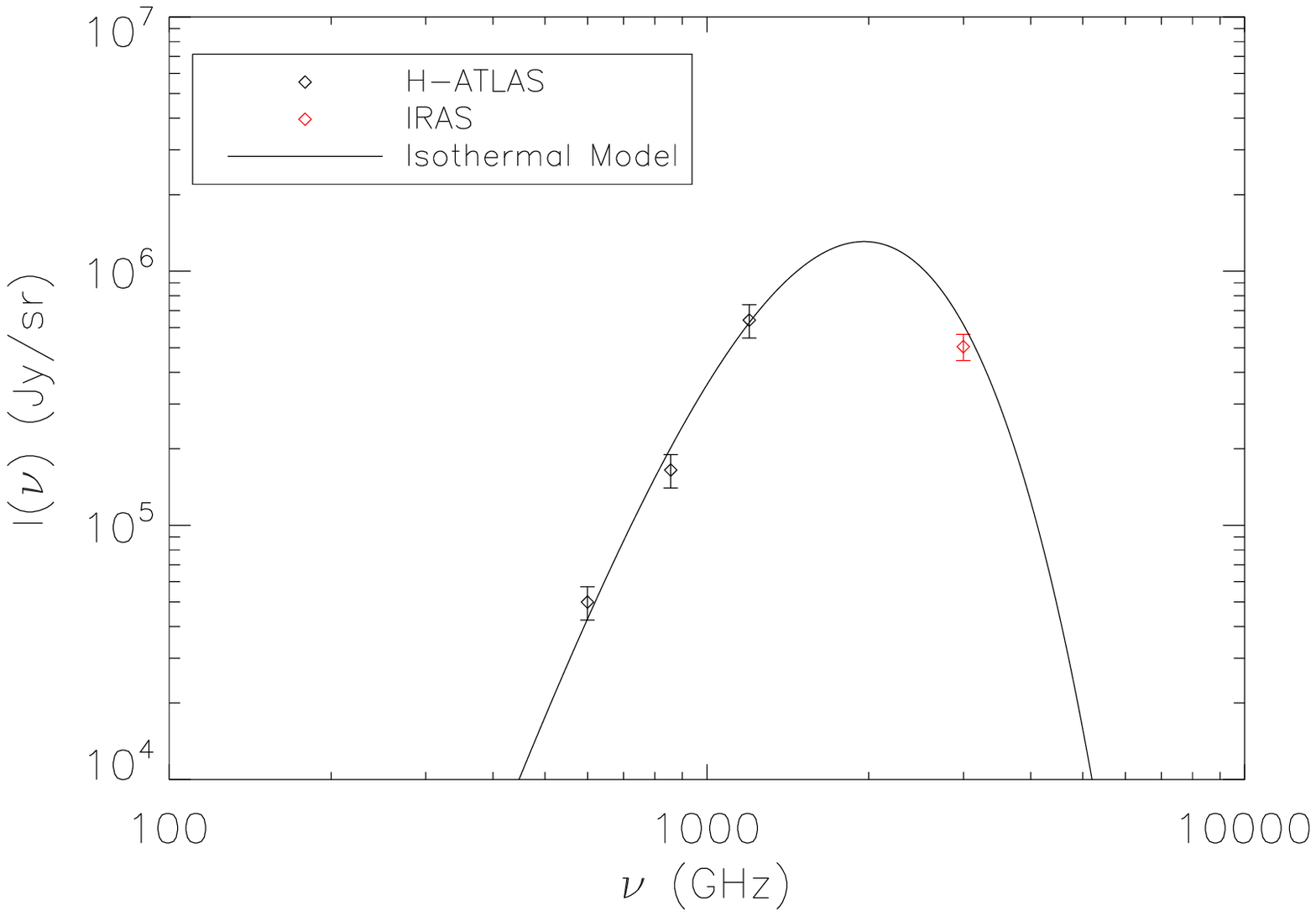,scale=0.45}
\end{minipage}
\caption{Example SEDs showing SPIRE and {\it IRAS} 6$'$ pixel intensities and the best-fit isothermal model with $T_{\rm d}=20.1$ and $\beta=1.3$ (top), $T_{\rm d}=18.2$ and $\beta=2.1$ (middle),
and $T_{\rm d}=13.1$ and $\beta=4.0$ (bottom).
}
\end{figure}

Instead of simply using the best-fit $T_{\rm d}$ and $\beta$ values and their variances when doing a fit to
the two forms of $\beta(T_{\rm d})$ given above, we need a numerical method that also takes into account the full covariance between the two parameters at each of the pixels.
To achieve this we fit for the two parameters describing each of the relations between $\beta$ and $T_{\rm d}$ by making use of the full probability distributions captured by the 
MCMC chains. The procedure we use  is the same as that of Veneziani et al. (2010).

A basic summary of the approach is that we fit, for example $N$ and $\alpha$, by drawing random pairs of
$\beta$ and $T_{\rm d}$ by sampling their likelihood functions from the MCMC chains we had first generated by fitting the isothermal SED models
to individual pixel intensities; here again, we restrict the analysis to chains where $\beta$ and $T_{\rm d}$ are determined with relative errors better than 30\% for both parameters.

By using the full MCMC chains to sample $\beta$ and $T_{\rm d}$ directly we keep information related to the
full covariance and this takes into account the fact that $\beta$ and $T_{\rm d}$ are anti-correlated in each of the pixels that we use for this analysis. 
For each of the two forms of $\beta(T_{\rm d})$, we sample the chains by drawing 20,000 random pairs  of $\beta$ and $T_{\rm d}$; we established a sampling of 20,000 
is adequate by a series of
simulations using anti-correlated data points in the $\beta-T_{\rm d}$ diagram with errors consistent with Figure~4 and assuming random correlation coefficients of -0.3 to -0.8.
Through this fitting procedure, we extract the distribution functions of the
four parameters $N$, $\alpha$, $C$, and $x$ and these in return allow us to quote their best-fit values and errors.

We find  $[N, \alpha]$ of [$N=116\pm38$, $\alpha=-1.4 \pm 0.1$] and
$[C,x]$ of [$C=-0.36\pm 0.02$, $x=(5.1\pm0.1)\times10^{-2}$]. We show these two best-fit lines in Figure~4.
The two models have $\chi^2$ per degree-of-freedom values of 0.8 and 0.5, respectively, suggesting that the model of Dupac et al. (2003) is slightly preferred
over the other. With reduced $\chi^2$ values less than one, it is
 likely that we are also overestimating our overall error budget, especially with the 15\% flux calibration uncertainty. In Swinyard et al.
(2010) the calibration error for SPIRE data is stated with an additional safety margin and the likely error is between 5\% and 10\%.
For comparison, D\'esert et al. (2009) found
[$N=11.5 \pm 3.8$, $\alpha=-0.66 \pm 0.05$], while Dupac et al. (2003) found
 $C=0.40 \pm 0.02$ and $x=(7.9 \pm 0.5)\times10^{-3}$.
These two lines, as well as the best-fit line of Veneziani et al. (2010), are shown in Figure~4 for comparison.
The Veneziani et al. (2010) measurements involve 7 high density clouds in the BOOMERanG-2003 CMB field and their measurements are consistent with 
prior works, except for one cloud with a low dust temperature of ($6.5 \pm 2.6$)K and $\beta$ of $5.1\pm 1.8$.

While we find $\beta$ values as high as 4 to 5, most of the values are between 1 and 3. We show three example SEDs in Figure~6 spanning low, mid, and high values of both $T_{\rm d}$ and $\beta$. Even if we constrain the study of $\beta(T_{\rm d})$ relation to the range of $1 < \beta < 3$, we still find non-zero
values for the four parameters  $N$, $\alpha$, $C$, and $x$ consistent with above values, 
suggesting that there is an intrinsic anti-correlation and not driven 
by the few high $\beta$ points. 

Figure~4 demonstrates two interesting scientific results: (i) we find an underlying relation between $T_{\rm d}$ and $\beta$ that 
cannot be due to an anti-correlation induced
by noise when fitting the SED form to pixel intensities and (ii)  
 we find a higher value of $\beta$ at low dust temperatures compared to the values sugested by previous relations in the literature.
Our result shows that the anti-correlation also exists for low-intensity Galactic cirrus and is not limited to high density clumps and molecular
clouds that were previously studied. It is likely that the $T_{\rm d}$ vs. $\beta$ relation captures different physical and chemical properties of the dust grains, including the size distribution and the interstellar radiation field that heat the dust. 
There is also a possibility that this anti-correlation results from line of sight projection of different dust temperature
components (Shetty et al. 2009b). With the field at a Galactic latitude of 30 degrees, such overlap is likely to be small, but perhaps not completely negligible. 
Unfortunately we do not have a way to constrain the line-of-sight projection due to the lack of 
distance information. Also, studies on the $\beta$ and $T_{\rm d}$ relation are 
so far limited to handful of fields
and more work is clearly desirable. We note that laboratory measurements have suggested the possibility of such an anticorrelation for
certain types of dust grains (e.g., Agladze et al. 1996; Mennella et al. 1998; Boudet et al. 2005). This 
has been explained as due to quantum physics effects on the amorphous grains, such as due to
two-phonon processing and tunneling effects between ground states of multi-level systems. Whether the relation we have observed is due to averaging different values of temperature along the line-of-sight
or due to an intrinsic property of dust is something that will remain uncertain.

Related to the observation (ii) outlined above, the measurements we report here are primarily dominated by the diffuse cirrus emission over the whole SDP area.
The Dupac et al. (2003) relation was for a large sample of molecular clouds in the Galaxy while the D\'esert et al. (2008) measurements
involve a sample of cold clumps detected as point sources in the Archeops CMB experiment. The dust size distribution is expected to be different in denser
regions compared to that in diffuse cirrus as the small grains are expected to coagulate into large aggregates. The diffuse cirrus is likely
dominated by small grains and this difference could be captured in terms of different values of $\beta$ for a given $T_{\rm d}$. The expectation is that denser regions would
show smaller values of $\beta$ (e.g., Ossenkopf \& Henning 1994), consistent with Figure~4. 
Once {\em Herschel} imaging data have been obtained for more of the H-ATLAS areas it will be interesting to study the $T_{\rm d}$
and $\beta$ relation for a variety of source structures, 
from dense cores and clumps in our Galaxy to extragalactic sources to diffuse Galactic cirrus emission in order to
establish how the relation changes with the environment. Both the Hi-GAL survey with {\it Herschel} (Molinari et al. 2010) 
and Planck can make important contributions to this topic in the future.

\begin{figure*}
\label{fig:map250}
\epsfig{file=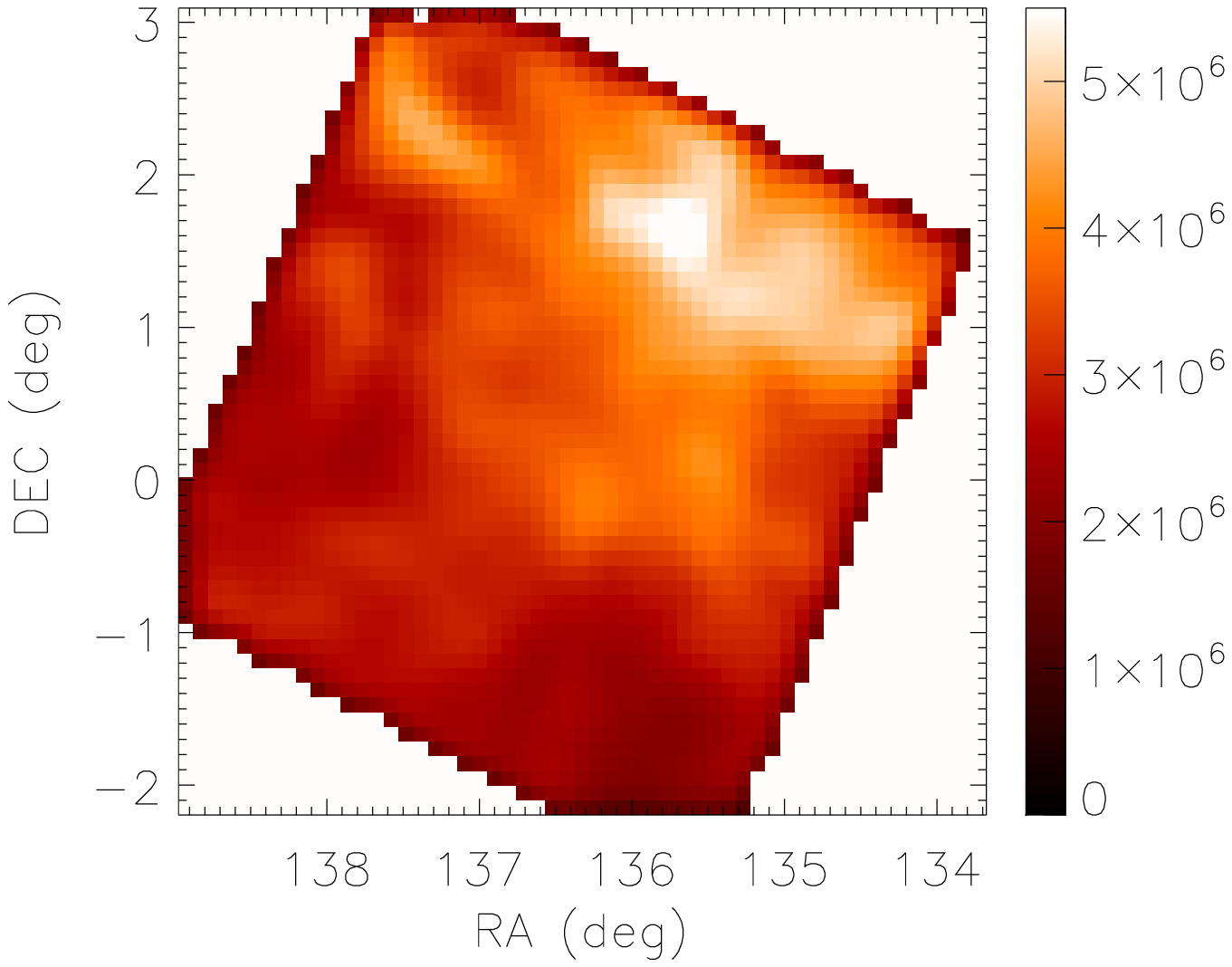,scale=0.3}
\epsfig{file=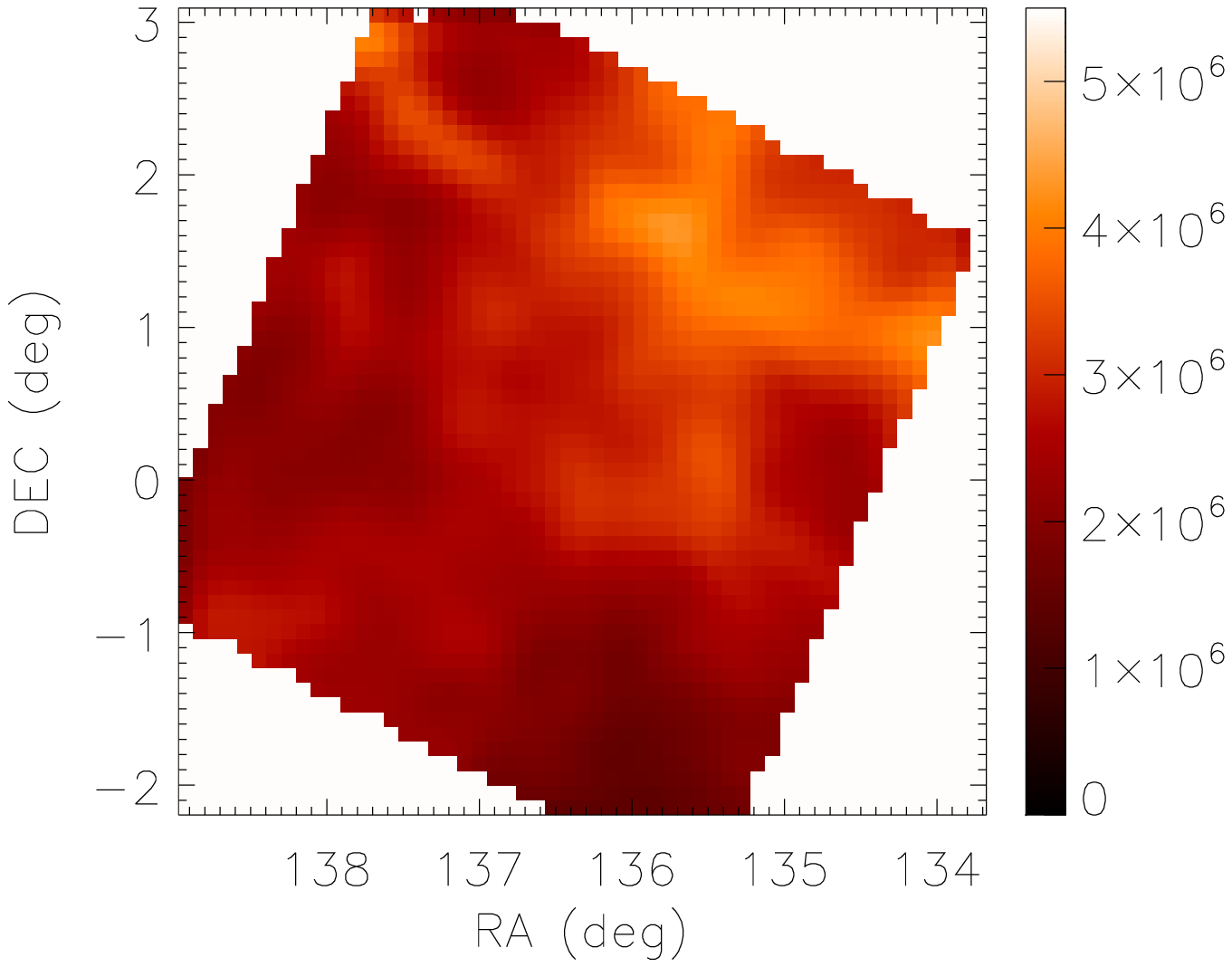,scale=0.3}
\epsfig{file=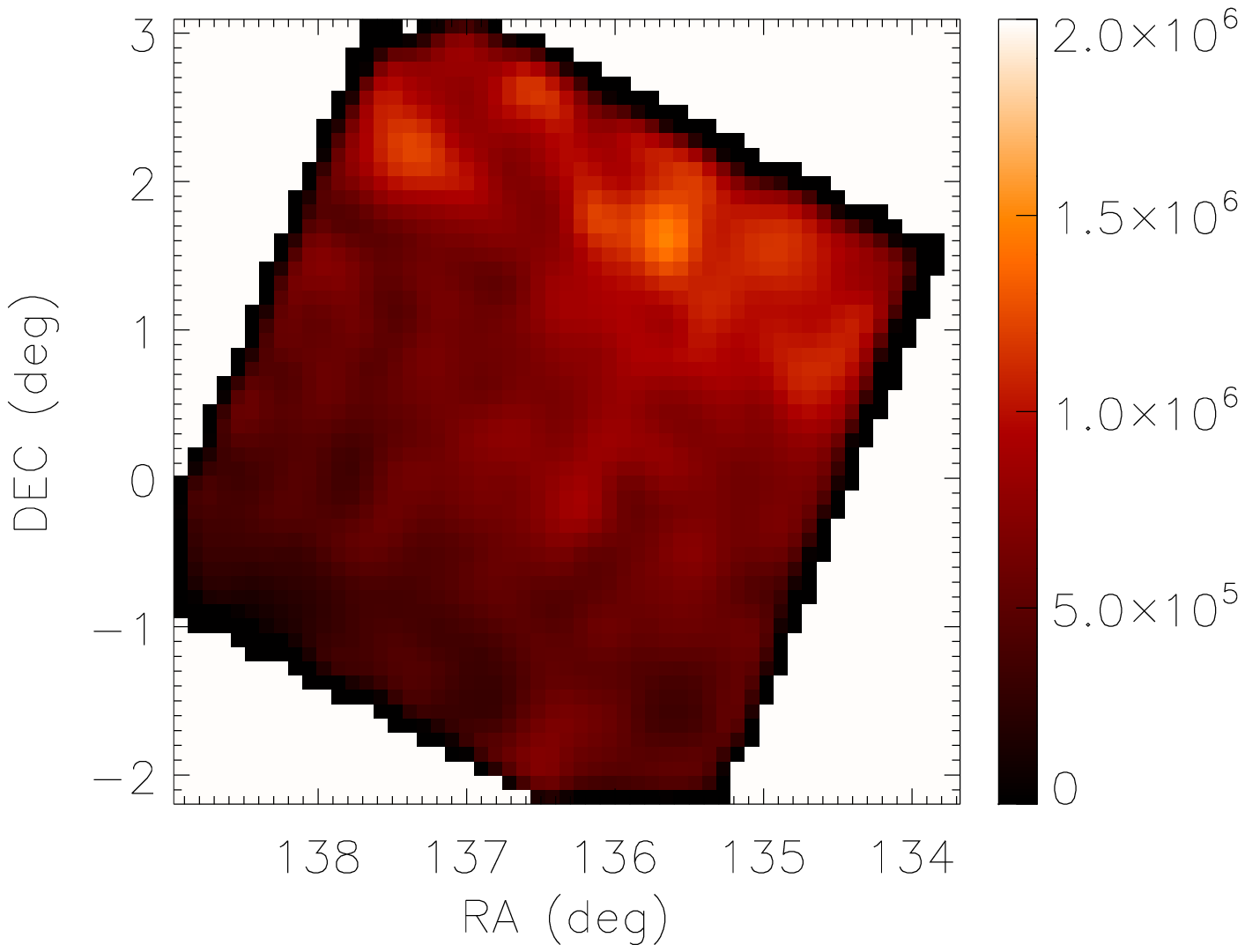,scale=0.3}
\caption{\textit{Left}: SPIRE 250 $\mu$m map of H-ATLAS SDP field at the 6$'$ pixel resolution used for studying $\beta$ and $T_{\rm d}$ over the whole map  (Section~4.4).
\textit{Center}: A map of the same field extracted from the Schlegel et al. (1998) dust map with Finkbeiner et al. (1999) model-8 frequency scaling to
obtain 250 $\mu$m intensities at the same 6$'$ pixel resolution. 
\textit{Right}: A map showing the difference between the observed data and the extrapolation from the 
Schlegel et al. (1998) dust map with Finkbeiner et al. (1999) model-8 frequency scaling. 
All three maps are in units of Jy/sr. To highlight the difference map, the third map has a shorter color stretch.}
\end{figure*}

\begin{figure*}\label{fig:betatempm}
\begin{minipage}{100mm}
\epsfig{file=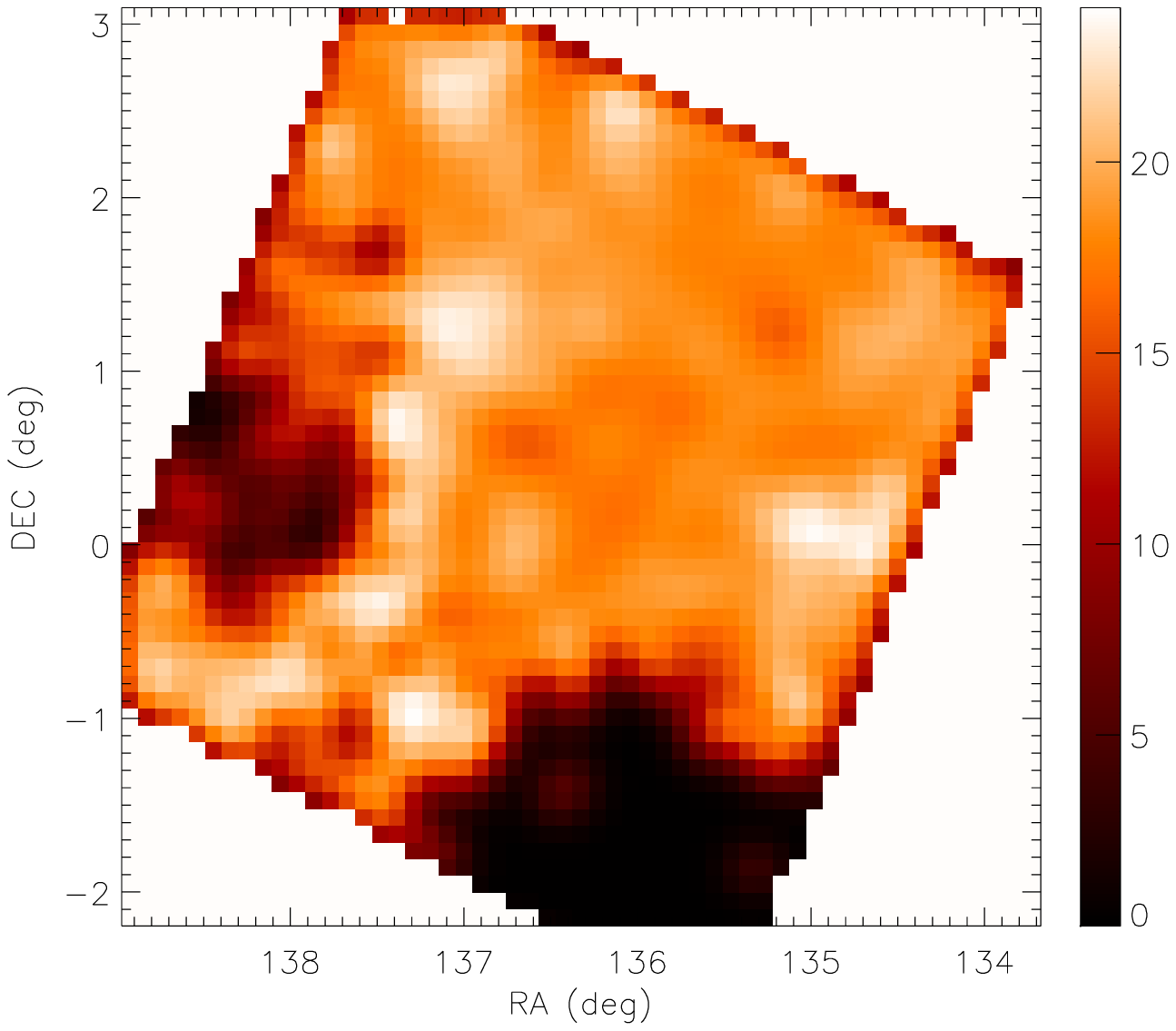,scale=0.6}
\epsfig{file=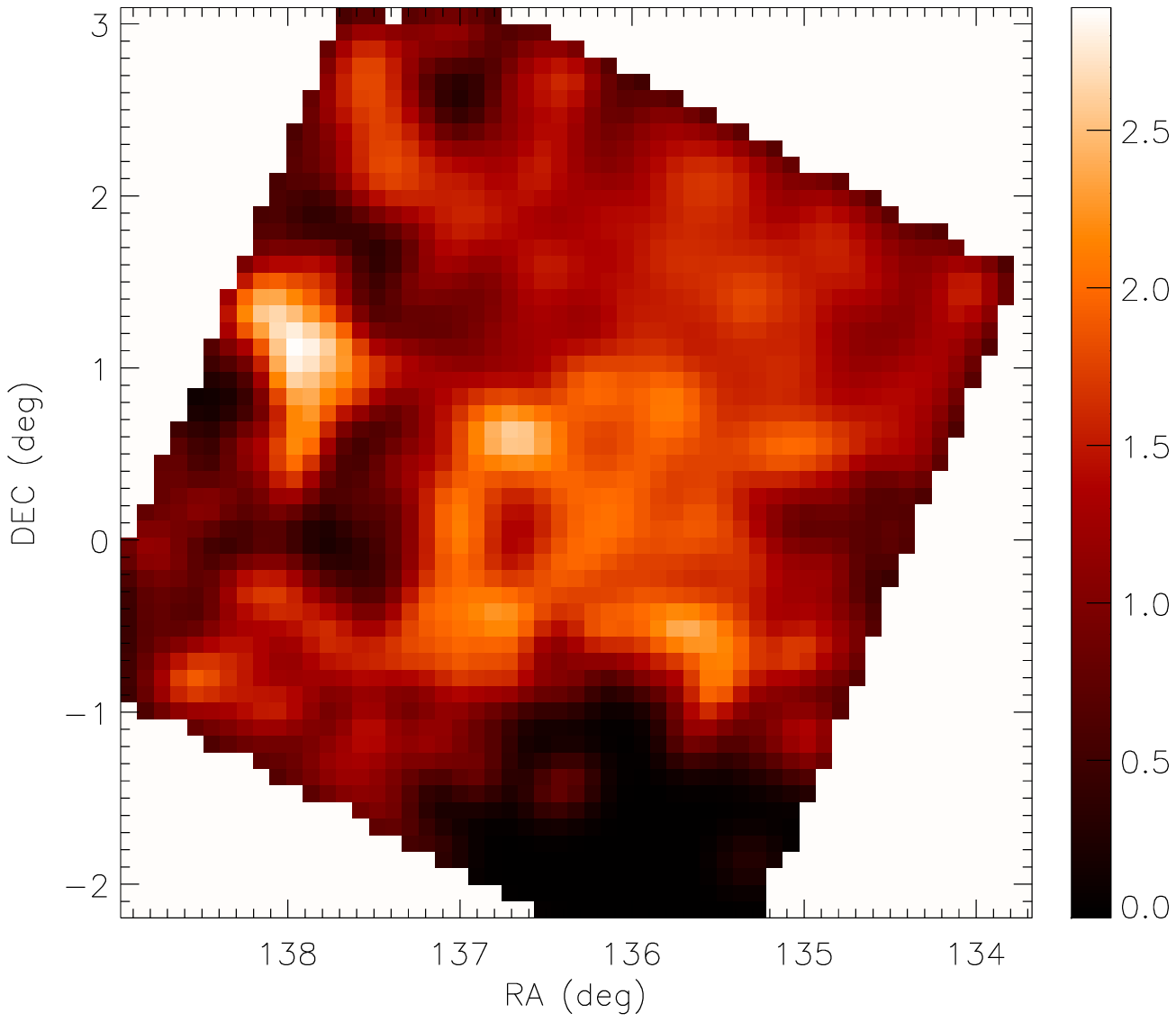,scale=0.6}
\end{minipage}\begin{minipage}{100mm}
\epsfig{file=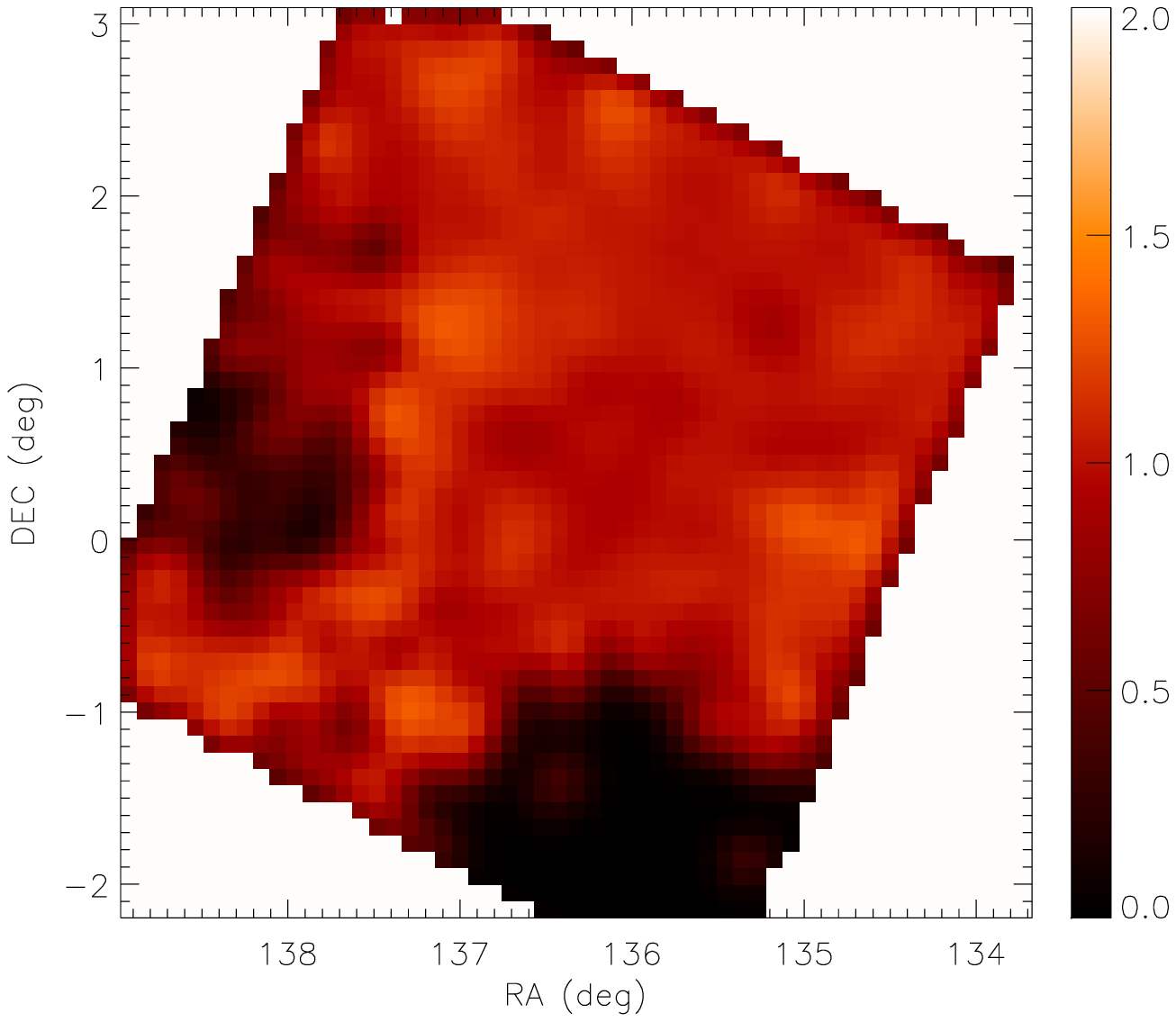,scale=0.6}
\epsfig{file=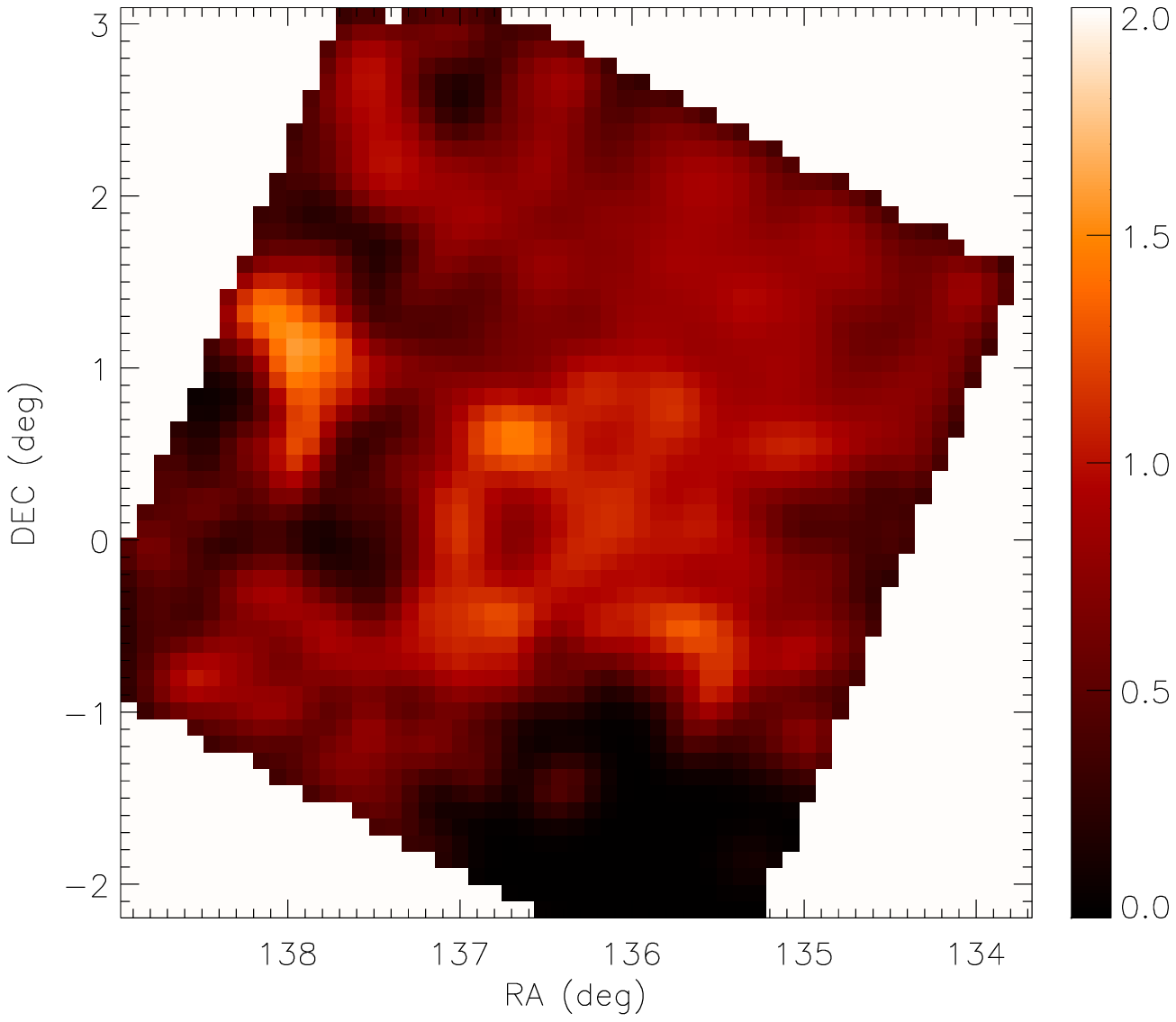,scale=0.6}
\end{minipage}
\caption{ \textbf{Top:} \textit{Left} - Map of the dust temperature: best fit values are shown. Units are degrees Kelvin. 
Pixels with $T_{\rm d}<5$\,K are dominated by noise and should not be considered as an accurate measurement. Figure~5 shows the
subselection where $T_{\rm d}$ and $\beta$ are determined with a relative accuracy better than 30\%.
\textit{Right} -  Ratio map between data temperature map and  a map of the dust temperature 
by reanalyzing the Schlegel et al. (1998; SFD) dust 
map  for the same region scaled by model-8 of the Finkbeiner et al. (1999) two temperature description 
and refitted with our isothermal SED following the same procedure
as data.  \textbf{Bottom :} \textit{Left} - Map of spectral emissivity parameters $\beta$, best fit values are shown. \textit{Right} - Map of the ratio between data spectral emissivity map and the pectral emissivity map derived from the SFD model dust map similar to dust temperature.}
\end{figure*}

\begin{figure}
\includegraphics[scale=0.5]{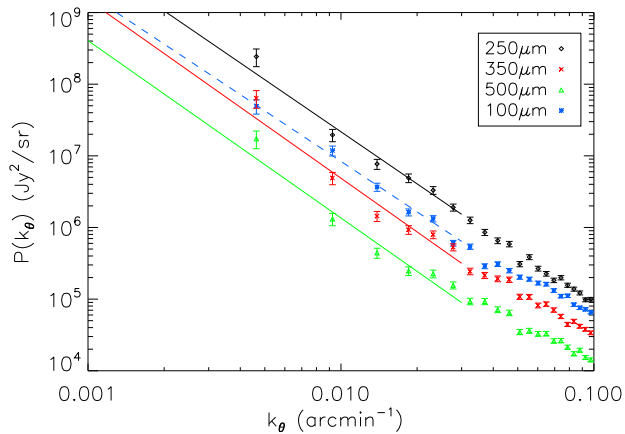} 
\caption{Angular power spectrum of the H-ATLAS field associated with Galactic cirrus. The power-law model fits are considered down to  $k$ of 0.03 arcmin$^{-1}$ to avoid
the cirrus measurements with extragalactic background fluctuations that are expected to peak at $k \sim 0.1$ arcmin$^{-1}$.}
\label{fig:fitpower}
\end{figure}

\subsection{Comparison to a dust model} 

We compare our  $\beta$ and $T_{d}$ maps with analogous maps obtained through model 8 of Finkbeiner et al. (1999) following the dust model of
Schlegel et al. (1998; SFD hereafter). The SFD dust map is produced  by combining
100 $\mu$m {\it IRAS} and 240 $\mu$m COBE/DIRBE data and is an all-sky map of sub-millimeter and microwave emission of the diffuse interstellar dust.
The model 8 of Finkbeiner et al. (1999) involves two dust temperature components at 9.2 and 16.2\,K with $\beta=1.67$ and 2.70, respectively (see the two points in Figure~4).
The SFD dust map with a frequency scaling such as model 8 is heavily utilized by the CMB experimental community both in planning and quantifying the 
Galactic dust contamination in CMB anisotropy 
measurements. At tens of degree angular scales and at frequencies above 90 GHz, Galactic dust is expected to be the dominant foreground
contamination, especially for polarization measurements of the CMB (e.g., Dunkley et al. 2009).

In order to compare the measurements in the SDP field to predictions from the SFD map combined with Finkbeiner et al. (1999) frequency scaling for dust emission, 
we make a new set of maps at 100, 250, 350, and 500 $\mu$m using the SFD dust map and overlapping with the H-ATLAS SDP field (Figure~7 shows a 
comparison of maps at 250 $\mu$m). We analyzed
these four simulated maps by applying the same procedure as we used for extracting  spectral emissivity parameter and dust temperature with SPIRE and {\it IRAS} real maps.
We also include the SPIRE noise making use of the SPIRE noise maps of the field generated from the data.
We obtain two maps from the SFD dust map, one connected to the spectral emissivity $\beta$ and one related to the dust 
temperature $T_{\rm d}$. 

We show the ratio map between data and SFD model results for both spectral emissivity and dust 
temperature in Figure~8 (right panels), while the left panels show our measurements directly on  SPIRE data.
The two component dust model of Finkbeiner et al. (1999)  only captures a limited range of  dust temperature and $\beta$.
The model involves a low-dust temperature component at 9.2\,K with $\beta=1.67$;  SPIRE maps spanning out to 500 $\mu$m are not strongly sensitive to such a cold
dust component though such a cold component primarily impacts the mm-wave data. We find that the mid values are safely produced by the SFD model, but is not a reliable description for
 regions that have either low or high temperatures. Such regions have either low or high $\beta$ values due to the anti-correlation between the two parameters.
Wide-field imaging with {\em Herschel}, such as the existing H-ATLAS and the proposed {\em Herschel}-SPIRE Legacy Survey (Cooray et al. 2010) and Planck, will provide necessary information to
improve the Galactic dust map and the associated frequency scaling model as a function of the sky position. 
While in this work we only considered the 14 deg$^2$ SDP field, the SGP and NGP portions of H-ATLAS each cover
about 200 deg$^2$, will significantly improve the IRAS-based dust model of our Galaxy, especially at Galactic latitudes probed by CMB experiments. 
In a future paper we will return to a further analysis on the improvements necessary for the dust model with data in those wide fields.

\section{Cirrus Power Spectrum}

Given that SPIRE is capable of mapping the diffuse emission at large angular scales, we also study the angular power spectrum related to cirrus emission
for each of the wavelength bands. For this measurement we  keep the maps at the original pixel scale
(Section~\ref{sec:datasets}) and compute the power spectrum of the intensity in maps masked for detected sources. With 1\% to 5\% of the pixels masked,
we found the mode coupling introduced by the source mask to be negligible.

We show our measurements in Figure~9. At $k > 0.1$ arcmin$^{-1}$ the effects of the beam transfer function become important but we do not
make a correction for the beam here as we are mostly interested in the power spectrum at degree angular scales, where the transfer function related to the SPIRE beam is
effectively one (Martin et al. 2010; Miville-Desch\^enes et al. 2010; Amblard et al. 2010). In addition to the beam, there is also the map-making transfer function
associated with any filtering employed during the map-making process (Pascale et al. 2010). At the angular scales of interest this transfer function is also consistent with
one. 

We assume a power-spectrum of the form $P(k)=P_{0}(k/k_0)^{\gamma}$ to describe the measurements and take
$k_0=0.01$ arcmin$^{-1}$ to be consistent with previous studies.
Using data out to $k < 0.03$ arcmin$^{-1}$, to avoid contamination with fluctuations associated with the extragalactic background, we  find $\gamma=-2.4 \pm 0.1$ at 100 $\mu$m with IRAS
and $-2.6\pm0.2$ at each of the SPIRE bands at 250, 350, and 500 $\mu$m. The measured values of the fluctuation amplitudes are $P_0=(8.3\pm0.7)\times10^6$, ($2.2\pm0.2$)$\times10^7$,
 $(4.8\pm0.5)\times10^6$, $(1.3\pm0.1)\times10^6$ at 100, 250, 350, and 500 $\mu$m respectively (in units of Jy$^2$/sr).
The normalization at 100 arcmin$^{-1}$ angular scale we find for {\it IRAS} 100 $\mu$m map is fully consistent with the 
relation between $P(k=0.01)$ and the mean 100 $\mu$m intensity in  Miville-Desch\^enes et al. (2007; see their Figure~4),
given the {\it IRAS} mean intensity of 1.77 MJy sr$^{-1}$ in the SDP field.
The power-law slope $\gamma$ we find is somewhat lower than  measurements for the slope in the literature 
(e.g., Miville-Desch\^enes et al.  2010 with a slope if $-2.7 \pm 0.1$, Martin et al. 2010 with slopes of $-2.74 \pm 0.03$ and $-2.81 \pm 0.03$ in two separate fields), but consistent with the analysis in Roy et al. 2010 (with a slope of $-2.6 \pm 0.07$ at 250 $\mu$m). All these measurements are consistent with each other given the overall  uncertainties.
It could also be that our slope is lowered by tens of percent level due to fluctuations associated with the extragalactic background. If 
we constrain to $k < 0.01$ arcmin$^{-1}$, keeping only two data points, we do find a higher slope closer to -2.8 to -2.9 but with larger uncertainty ($\pm 0.4$).

The $\sqrt{k^2P(k)/2\pi}$ captures the rms fluctuations over the whole SDP area arising from Galactic cirrus and the extragalactic background at a given value of the wavenumber.
At large angular scales, the fluctuations generated by the extragalactic sources (mainly the sources contributing the background confusion noise) are subdominant with 
values of order $10^4$ to $10^5$  Jy$^2$/sr in  $P(k)$  when $k < 0.01$ arcmin$^{-1}$ (say at 250 $\mu$m; Amblard et al. 2010). For comparison, in  Figure~9
we find Galactic cirrus fluctuations at the level of 10$^7$ to 10$^9$  Jy$^2$/sr. Thus, at $k=0.01$ arcmin$^{-1}$ we can safely assume that all of the fluctuations
we have measured arise from Galactic cirrus. Taking the $\sqrt{(P_0)}$ values we can make an independent estimate of the dust temperature and $\beta$ similar to the
analysis of Martin et al. (2010). We find $T_{\rm d}=20.1 \pm 0.9$\,K and $\beta=1.3 \pm 0.2$,
consistent with the same two quantities we obtained in Section 4.2 by cross-correlating SPIRE maps with IRAS. 

\section{Summary and Conclusions}

In this paper, we have studied the Galactic dust SED and the angular power spectrum of dust fluctuations in the
14 deg$^2$  Science Demonstration Phase (SDP) field  of H-ATLAS. 
By correlating  the SPIRE 250, 350 and 500 $\mu$m and {\it IRAS} 100 $\mu$m maps to extract the sub-mm color terms
 of SPIRE maps relative to the {\it IRAS} 100 $\mu$m map of the SDP field,
we find the average dust temperature over the whole field to be $19.0 \pm 2.4$\,K with the spectral emissivity parameter taking a value of $1.4 \pm 0.4$.
We find all of the bright cirrus regions to have dust temperatures $T_{\rm d}$ over a narrow range of 17.4 to 18.3\,K ($\pm 2.5$\,K),
with a spectral emissivity parameter  $\beta$ ranging from  1.4 to 1.9 ($\pm0.5$). 
Similar to previous studies, we find an anti-correlation between $T_{\rm d}$ and $\beta$;
when described by a power-law with $\beta=AT_{\rm d}^\alpha$, we find $A=116\pm38$ and $\alpha=-1.4 \pm 0.1$ while 
a relation of the form $\beta = (C+xT_{\rm d})^{-1}$ is also consistent with data with $C=-0.36\pm 0.02$ and $x=(5.1\pm0.1)\times10^{-2}$.
The observed inverse relation between $T_{\rm d}$ and $\beta$ is stronger than the previous suggestions in the literature
and we have suggested the possibility that this stronger anti-correlation may be due to the fact that we study primarily diffuse cirrus while previous studies 
involved high density environments such as molecular clouds and cold clumps.
We also make an independent estimate of the dust temperature and  the spectral emissivity parameter, when averaged over the whole field, through the frequency scaling
of the rms amplitude of dust fluctuation power spectrum. At 100 arcminute angular scales, we obtain
$T_{\rm d}=20.1 \pm 0.9$\,K and $\beta=1.3 \pm 0.2$, consistent with previous estimates.
The cirrus fluctuations power  spectrum is consistent with a power-law at 100, 250, 350 and 500 $\mu$m 
with a power-law spectral index of $-2.6\pm0.2$ from 1 to 200 arcminute angular scales.

After we completed this paper, we became aware of a similar study involving the $\beta$ and $T_{\rm d}$ in the first two fields covered by the Hi-GAL survey (Paradis et al. 2010).
These authors also find an anti-correlation between the two parameters, but the mean relation is distinctively different between the two fields at Galactic longitudes of 30 and 50 degrees, with
both on the Galactic plane.
When compared to the H-ATLAS SDP field, these two fields have mean intensities that are a factor of 200 larger at the level of  1000 MJy sr$^{-1}$. 
Interestingly the $\beta(T_{\rm d})$ relation they find for the field at $l=30$ degrees is consistent with the relation we report here
when extrapolating their relation that was determined over the range of $1.5 < \beta < 2.5$ and $18 < T_{\rm d}/{\rm K} < 25$ to lower 14\,K dust temperatures we find in
some of the pixels in our field with $\beta \sim 4$. It could very well be that the dust properties are far more complex than the simple isothermal models we have 
considered and a variety of effects may be contributing to the observed anti-correlation.  Further studies making use of wide area maps are clearly warranted.

\section*{Acknowledgments}

The {\em Herschel}-ATLAS is a project with {\em Herschel}, which is an ESA space observatory with science instruments provided by 
European-led Principal Investigator consortia and with important participation from NASA. The H-ATLAS website is http://www.h-atlas.org/
Amblard, Bracco, Cooray, and Serra acknowledge support from NASA funds for US participants in 
{\em Herschel} through JPL.

\end{document}